\documentclass[11pt]{article}
\usepackage{amssymb,amsmath}
\usepackage{graphicx}
\usepackage[english]{babel}

\newtheorem{DE}{Definition}[section]

\newcommand {\sm} {\setminus}
\usepackage{latexsym} %Des symboles de maths en plus.
\usepackage{theorem} %pour changer les styles dans les theoremes.
\usepackage{graphicx}

\newcommand{\qed}{\relax\ifmmode\hskip2em\Box\else\unskip\nobreak\hfill$\Box$\fi}

\newtheorem{theorem}[DE]{Theorem}
\newtheorem{lemma}[DE]{Lemma}
\newtheorem{conjecture}[DE]{Conjecture}

{\theoremstyle{break}\theorembodyfont{\rmfamily}}
{\theoremstyle{break}\theorembodyfont{\rmfamily}}

\newcounter{claim}
\newenvironment{proof}[1][]%
	{\noindent {\setcounter{claim}{0}\sc proof --- }{#1}{}}{\qed\vspace{2ex}}
	{\refstepcounter{claim}\vspace{1ex}\noindent {(\it\arabic{claim}) {#1}{}}\it}{\vspace{1ex}}
	{\noindent {}{#1}{}}{ This proves~(\arabic{claim}).\vspace{1ex}}

\begin{document}

\title{$\chi$-bounds, operations and chords}

\author{Lan Anh Pham\thanks{Department of Mathematics and Mathematical Statistics,
Ume\aa\
  University, Sweden. e-mail: lan.pham@umu.se}~~and Nicolas
  Trotignon\thanks{CNRS, LIP, ENS de Lyon. Partially supported by ANR project Stint
    under reference ANR-13-BS02-0007 and by the LABEX MILYON
    (ANR-10-LABX-0070) of Universit\'e de Lyon, within the program
    ``Investissements d'Avenir'' (ANR-11-IDEX-0007) operated by the
    French National Research Agency (ANR).  Also Universit\'e Lyon~1,
    universit\'e  de Lyon. e-mail:
  nicolas.trotignon@ens-lyon.fr}}

\maketitle

\begin{abstract}
  A \emph{long unichord} in a graph is an edge that is the unique
  chord of some cycle of length at least 5.  A graph is
  \emph{long-unichord-free} if it does not contain any long-unichord.
  We prove a structure theorem for long-unichord-free graph. We give
  an $O(n^4m)$-time algorithm to recognize them. We show that any
  long-unichord-free graph $G$ can be colored with at most $O(\omega^3)$
  colors, where $\omega$ is the maximum number of pairwise adjacent
  vertices in $G$.

\end{abstract}



\section{Introduction}
\label{sec:intro}

In this article, all graphs are finite and simple.  We denote by
$\chi(G)$ the \emph{chromatic number} of a graph $G$, that is the
minimum number of colors needed to give a color to each vertex in such
a way that any  two adjacent vertices receive different colors.  We denote by
$\omega(G)$ the maximum size of a set of pairwise adjacent vertices
(that we call a \emph{clique}).  It is clear that for every graph,
$\chi(G) \geq \omega(G)$, while the converse inequality is false in
general (the smallest example is the chordless cycle on five
vertices).

Let $f$ be any function from $\mathbb R$ to $\mathbb R$.  A graph $G$
is \emph{$\chi$-bounded by $f$} if every induced subgraph $H$ of $G$
satisfies $\chi(H) \leq f(\omega(H))$.  This notion first appeared in
an article of Gy\'arf\'as~\cite{gyarfas:perfect}.   A class of
graphs is \emph{$\chi$-bounded} if for some function $f$, every graph
of the class is $\chi$-bounded by $f$.  It is well known that the
class of all graphs is not $\chi$-bounded, this follows from the
existence of graphs with $\omega=2$ and arbitralily large chromatic
number, see for instance~\cite{Zyk49}.

Graphs that are $\chi$-bounded by the indentity function are known as
\emph{perfect graphs}.  They were the object of much research
(see~\cite{nicolas:perfect} for a survey), and the notion of
$\chi$-boundedness was invented to try to have some insight on them.
In his seminal paper, Gy\'arf\'as~\cite{gyarfas:perfect} made many
conjectures, and some of them were claiming that excluding chordless
cycles with various constraints on their length should lead to
$\chi$-bounded classes.  Recently, much progress has been reported
toward these conjectures, see for instance~\cite{ChudnovskySS16}.

In this paper, we focus on excluding cycles with contraints on their
chords.  A~\emph{unichord} in a graph is an edge that is the unique
chord of some cycle (note that the cycle has length at least 4 because
of the chord).  A graph is \emph{unichord-free} if it does not contain
any unichord.  A \emph{long unichord} in a graph is an edge that is
the unique chord of some cycle of length at least~5. The \emph{house}
is the graph on five vertices $a, b, c, d, e$ with the following
edges: $ab, bc, cd, da, ea, eb$ (so the house is the smallest graph
that contains a long unichord).  A \emph{house*} is any graph obtained
from the house by repeatedly subdividing edges.  A graph is
\emph{house*-free} if it does not contain any house* as an induced
subgraph.

It is straightforward to check that long-unichord-free graphs form a
generalisation of unichord-free graphs and of house*-free graphs.
They also form a generalisation of the classical class of chordal
graphs (a graph is \emph{chordal} if it contains no chordless cycle of
length at least~4).  A classical result states that chordal graphs
are perfect (equivalently, they are $\chi$-bounded by the indentity
function).  In~\cite{nicolas.kristina:one}, it is proved that
unichord-free graphs are $\chi$-bounded by the function
$f(x) = \max(3, x)$, and in~\cite{penev:amalgam}, it is proved that
house*-free graphs are $\chi$-bounded by some exponential function.
We generalise these theorems and we provide a better bound for the
last one by showing that long-unichord-free graphs are $\chi$-bounded
by a polynomial function of degree~3, namely $f_3$, to be defined later: 

{
\renewcommand{\theDE}{\ref{the:cluc}}

\begin{theorem}
  Long-unichord-free graphs are $\chi$-bounded by $f_3$ (in
  particular, by a polynomial of degree~3).
\end{theorem}
\addtocounter{DE}{-1}
}

Our proof relies on a decomposition theorem that is easily obtained
from~\cite{nicolas.kristina:one} and~\cite{conforti.c.k.v:capfree}
(again, we postpone the definitions).

{
\renewcommand{\theDE}{\ref{th:MainDec}}

\begin{theorem}
  Let $G$ be a connected long-unichord-free graph.  Then either:
  \begin{itemize}
  \item $G$ is an induced subgraph of the Petersen graph;
  \item $G$ is an induced subgraph of the Heawood graph;
  \item $G$ is chordal;
  \item $G$ is bipartite and one side of the bipartition is made of
    vertices of degree at most~2;
  \item $G$ has a universal vertex;
  \item $G$ has a  cutvertex;
  \item $G$ has an amalgam;
  \item $G$ has proper 2-cutset.
  \end{itemize}
\end{theorem}
\addtocounter{DE}{-1}
}

It must be stressed that applying decomposition theorems to prove
$\chi$-boundedness is not at all straightforward.  There are several
papers dealing with the following question: when a prescribed
operation is applied repeatedly to some graphs from a class
$\chi$-bounded by $f$, is the larger class that is obtained
$\chi$-bounded by a possibly different function $g$?  This has been
answered positively for several operations, most notably for one that
we use in our decomposition theorem, the so called amalgam operation
(to be defined in the next section), see~\cite{penev:amalgam}.  But
the theorem from~\cite{penev:amalgam} is not enough for our purpose
because it leads to an exponential $\chi$-bounding function.  Here, to
obtain $\chi$-boundedness we prove a stronger property for the sake of
induction, roughly we find graphs with a special structure that
intersect all inclusion-wise maximal cliques of the graph to be
colored, and we apply this procedure inductively on what remains
(where the maximum clique is smaller).  

Our decomposition theorem turns out to be a complete structural
description of long-unichord-free graphs: it tells how all
long-unichord-free graphs can be constructed from simple pieces.  As a
byproduct of this description we obtain the following: 

{
\renewcommand{\theDE}{\ref{th:rec}}
\begin{theorem}
  Deciding whether an input graph $G$ has a long-unichord can be performed
  in time $O(n^4m^2)$ (where $n = |V(G)|$ and $m=|E(G)|$). 
\end{theorem}

\addtocounter{DE}{-1}
}

This answers an open question mentioned in~\cite{nicolas.kristina:one}
(where a similar algorithm is given for unichord-free graphs).  It
should be pointed out that in~\cite{nicolas.kristina:one}, a problem
of the very same flavour is proved to be NP-complete: deciding whether
a graph contains a cycle $C$ with a unique chord $uv$ such that $u$
and $v$ are at distance at least~4 along the cycle.  This shows that
naive attempts to obtain our recognition algorithm are likely to fail.

\subsection*{Outline of the paper}

In Section~\ref{sec:opkn}, we define all the decompositions and
operations on graphs that we need, and we survey several results about
how they preserve perfection and $\chi$-boundedness.

In Section~\ref{sec:subst}, we study a technique to prove that an
operation on graphs (namely, the substitution operation, everything is
defined in the next section) preserves $\chi$-boundedness.  This
technique is analogous to the one used in the proof of the replication
lemma of Lov\'asz (see~\cite{nicolas:perfect}). It consists in identifying a
particular subgraph that intersects all maximal cliques of a graph,
and in showing that the existence of such a subgraph is preserved by the operation.  Our
technique yields short proofs of known results and may provide good
bounds in some situations. For instance we prove that the closure of
3-colourable graphs under substitution is $\chi$-bounded by a
quadratic function, a seemingly new result.  Note that the results
from Section~\ref{sec:subst} are not used in the rest of the
article. They illustrate our method and are of independent interest.

In Section~\ref{sec:amalgam}, we apply a similar technique to a larger
set of operations (namely, we consider 1-joins, amalgams and proper
2-cutsets).  The price to pay for that is that the classes of graphs
where the technique can be applied are even more restricted. But
fortunately, it does not vanish to nothing as shown afterward.

In Section~\ref{sec:struc}, we prove the structure theorem for
long-unichord-free graphs. 

In Section~\ref{sec:cluc}, we apply the results of the previous
sections to prove that long-unichord-free graphs are $\chi$-bounded. 

In Section~\ref{sec:reco} we provide a polynomial time
algorithm to recognize long-unichord-free graphs, based on the
decomposition theorem.

Section~\ref{sec:open} is devoted to open questions.

\section{Operations and properties preserved by them}
\label{sec:opkn}

We now define several classical decompositions for graphs, that are
all partitions of the vertex-set with some structural constraints.  For
each of them, we explain how it enables us to obtain smaller graphs called
\emph{blocks of decomposition}, and how the decomposition can be
reversed into an operation that allows building a graph from smaller
pieces.

A vertex $x$ in a graph $G$ is \emph{complete} to
$A\subseteq V(G) \sm \{x\}$ if for all $y\in A$, $xy\in E(G)$.  We
also say that $x$ is \emph{$A$-complete}.  A set $A\subseteq V(G)$ is
\emph{complete} to a set $B\subseteq V(G)$ disjoint from $A$ if every
vertex of $A$ is $B$-complete.

A vertex $x$ in a graph $G$ is \emph{anticomplete} to
$A\subseteq V(G) \sm \{x\}$ if for all $y\in A$, $xy\notin E(G)$.  We
also say that $x$ is \emph{$A$-anticomplete}.  A set $A\subseteq V(G)$
is \emph{anticomplete} to a set $B\subseteq V(G)$ disjoint from $A$ if
every vertex of $A$ is $B$-anticomplete.

\subsection*{Gluing along a clique}
 
A (possibly empty) clique $K$ of a graph $G$ is a clique cutset of $G$
if there exists a partition $(X_1, K, X_2)$ of $V(G)$ such that
$X_1, X_2 \neq \emptyset$ and there are no edges of $G$ between $X_1$
and $X_2$.  We then say that $(X_1, K, X_2)$ is a \emph{split} for
this clique cutset, and that $G_1=G[X_1 \cup K]$ and
$G_2 = G[K \cup X_2]$ are the \emph{blocks of decomposition} of $G$
with respect to this split.  

Note that $G = G_1 \cup G_2$, and we say that $G$ is obtained from
$G_1$ and $G_2$ by \emph{gluing along a clique}.  This operation can
be performed for any pair of graphs $G_1, G_2$ such that
$G_1 \cap G_2$ is a clique.  If $K=\emptyset$, this operation can also
be refered to as \emph{disjoint union}.  When $|K|=1$, the operation
can be refered to as \emph{gluing along a vertex}, and the unique
vertex in $K$ is called a \emph{cutvertex}.

\subsection*{Substitutions}

A set $X$ of vertices of a graph $G$ is a \emph{homogeneous set} if
$|X| \geq 2$, $X\subsetneq V(G)$, and every vertex of $V(G) \sm X$ is
either complete or anticomplete to $X$. We then denote by $G/X$ the graph
obtained from $G$ by deleting $X$ and adding a vertex $v$ adjacent to
all $X$-complete vertices of $G$.  The graphs $G[X]$ and $G/X$ are the
\emph{blocks of decomposition of $G$} with respect to the 
homogeneous set~$X$. 

When $G$ is a graph on at least two vertices, $v$ is a vertex of $G$ and
$H$ is a graph on at least two vertices vertex-disjoint from $G$, then
the graph $G'$ obtained from $G$ by deleting $v$, adding $H$ and all
possible edges between vertices of $H$ and the neighbors of $v$ in $G$
is called the graph obtained from $G$ by \emph{substituting} $H$ for
$v$.  We also say that $G'$ is obtained from $G$ and $H$ by a
\emph{substitution}.  Observe that $V(H)$ is a homogeneous set of
$G'$.

\subsection*{1-join composition}

A \emph{1-join} of a graph $G$ is a partition of $V(G)$ into sets
$X_1$ and $X_2$ such that there exist sets $A_1, A_2$ satisfying:

\begin{itemize}
\item $\emptyset \neq A_1 \subseteq X_1$,
  $\emptyset \neq A_2 \subseteq X_2$;
\item $|X_1| \geq 2$ and $|X_2| \geq 2$;
\item there are all possible edges between $A_1$ and $A_2$;
\item there are no other edges between $X_1$ and $X_2$.
\end{itemize}

We say that $(X_1, X_2, A_1, A_2)$ is a \emph{split} of this 1-join.
For $i=1, 2$, the \emph{block of decomposition} $G_i$ with respect to
this split is the graph obtained from $G[X_i]$ by adding a
vertex $u_{3-i}$ complete to $A_i$.  

The operation that is the reverse of the 1-join decomposition is
defined as follows.  Start with two vertex-disjoint graphs $G_1$ and
$G_2$ on at least 3 vertices.  For some vertex $u_2$ (resp.\ $u_1$) of
$G_1$ (resp.\ $G_2$) such that $N_{G_1}(u_2)$ (resp.\ $N_{G_2}(u_1)$) is
non-empty, $G$ is obtained from the disjoint union of
$G_1 \setminus \{ u_2 \}$ and $G_2 \setminus \{ u_1 \}$ by adding all
possible edges between $N_{G_1}(u_2)$ and $N_{G_2}(u_1)$.  We say that $G$ is
obtained from $G_1$ and $G_2$ by a \emph{1-join composition}.

\subsection*{Amalgam composition} 

An \emph{amalgam} of a graph $G$ is a partition $(K, X_1, X_2)$ of
$V(G)$ such that $K$ is a (possibly empty) clique, $(X_1, X_2)$ is a
1-join of $G\sm K$ with a split $(X_1, X_2, A_1, A_2)$ and $K$ is complete
to $A_1\cup A_2$ (possibly, vertices of $K$ have neighbors in
$V(G) \sm (A_1 \cup A_2)$).

We say that $(X_1, X_2, A_1, A_2, K)$ is a \emph{split} of the amalgam
defined above.  For $i=1, 2$, the \emph{block of decomposition} $G_i$
with respect to this split is the graph obtained from $G[X_i \cup K]$
by adding a vertex $u_{3-i}$ complete to $A_i\cup K$.

The operation that is the reverse of the amalgam decomposition is
defined as follows.  Start with two graphs $G_1$ and $G_2$ whose
intersection forms a clique $K$ with $|K| \leq |V(G_1)|-3, |V(G_2)|-3$
and such that for $i=1, 2$ there is a vertex $u_{3-i}$ in
$V(G_i) \sm K$ whose neighborhood is $K \cup A_i$ where $A_i$ is
non-empty, disjoint from $K$ and $K$-complete.  Let $G$ be obtained
from the union of $G_1 \setminus \{ u_2 \}$ and
$G_2 \setminus \{ u_1 \}$ by adding all edges between $A_1$ and $A_2$.
We say that $G$ is obtained from $G_1$ and $G_2$ by an \emph{amalgam
  composition}.

The amalgam is obviously a generalisation of the 1-join.  If $A_1$ or
$A_2$ were allowed to be empty, it could be also be seen as a
generalisation of the clique cutset, but it is not (we keep this
distinction that might seem artificial, for historical reasons and
compatibility of definitions with previous papers).

If $X_1=A_1$ then $X_1$ is a homogeneous set of $G$.  Yet, formally
the amalgam is not a generalisation of the homogeneous set, because a
homogeneous set $X$ such that $|X|=|V(G)| - 1$ (which is allowed) does
not imply the presence of an amalgam. Setting $K=\emptyset$, $X_1=X$,
and $X_2 = V(G)\sm X$ does not work because then $|X_2| = 1$.  However,
it works for all homogeneous sets $X$ such that $|X| \leq |V(G)| - 2$.
This remark leads us to consider the following trivial decomposition
and lemma.

\subsection*{Adding a universal vertex}

A \emph{universal vertex} in a graph $G$ is a vertex $v$ complete to
$V(G)\sm\{v\}$.  Note that $V(G)\sm \{v\}$ is then a homogeneous set
of $G$ (that does not yield a 1-join or an amalgam as noted above).
From the discussion above, the following is trivial.

\begin{lemma}
  \label{l:easyS}
  If $G$ has a homogenous set, then either $G$ has a 1-join (and
  therefore an amalgam) or $G$ has a universal vertex.
\end{lemma}

The operation that is the reverse of ``having a universal vertex'' is
simply \emph{adding a universal vertex}, which means adding a vertex
$v$ to a graph $G$, and all possible edges between $v$ and $V(G)$.

\subsection*{Proper 2-cutset composition}

A \emph{proper 2-cutset} of a connected graph $G$ is a pair of
non-adjacent vertices $a, b$, such that $V(G)$ can be partitioned into
non-empty sets $X_1$, $X_2$ and $\{ a,b \}$ so that: $|X_1|\geq 2$,
$|X_2| \geq 2$; there are no edges between $X_1$ and $X_2$; and both
$G[X_1 \cup \{ a,b \}]$ and $G[X_2 \cup \{ a,b \}]$ contain a
path from $a$ to $b$.  We say that $(X_1, X_2, a, b)$ is a \emph{split} of this
proper 2-cutset.

For $i=1, 2$, the \emph{block of decomposition} $G_i$ with respect to
this split is the graph obtained from $G[X_i \cup \{a, b\}]$ by adding
a vertex $x_{3-i}$ complete to $\{a, b\}$.

The operation that is the reverse of the proper 2-cutset is defined as
follows.  Start with two graphs $G_1$ and $G_2$ whose intersection is
a pair of vertices $a, b$ non-adjacent in both $G_1$ and $G_2$, and
such that $a, b$ have a common neighbor $x_2$ in $G_1$, and a common
neighbor $x_1$ in $G_2$.  Suppose furthermore that for $i=1, 2$, there exists a
path from $a$ to $b$ in $G_i\sm x_{3-i}$.  Let $G$ be the union of
$G_1 \setminus \{ x_2 \}$ and $G_2 \setminus \{ x_1 \}$.  We say that
$G$ is obtained from $G_1$ and $G_2$ by a \emph{proper 2-cutset
  composition}.

\subsection*{Heredity of decompositions}

The next two lemmas are very easy to prove and we give them without
proofs. 

\begin{lemma}
  \label{l:hSubst}
  Suppose that $G$ is obtained from $G_1$ by substituting $G_2$ for
  $v$. If $G'$ is an induced subgraph of $G$, then either $G'$ is
  isomorphic to an induced subgraph of $G_1$, or $G'$ is an induced
  subgraph of $G_2$, or $G'$ is obtained from an induced subgraph of
  $G_1$ by substituting an induced subgraph of $G_2$ for $v$.
\end{lemma}

\begin{lemma}
  \label{l:lS}
  Suppose that $G$ is obtained from $G_1$ and $G_2$ by one of the
  operations from $S = \{$gluing
  along a clique, substitution, 1-join composition, amalgam
  composition, gluing along a proper 2-cutset$\}$.

  If $G'$ is an induced subgraph of $G$, then either $G'$ is
  isomorphic to an induced subgraph of $G_1$, or $G'$ is isomorphic to
  an induced subgraph of $G_2$, or $G'$ is obtained from an induced
  subgraph of $G_1$ and an induced subgraph of $G_2$ by an operation
  from $S$.
\end{lemma}

\subsection*{Properties preserved by the operations}
\nocite{berge.chvatal:topics}

Theorem~\ref{th:pop} below was proved by Gallai~\cite{gallai:triangule} (gluing
along a clique), Lov\'asz~\cite{lovasz:nh} (substitutions),
Cunningham~\cite{cunningham:1join} (1-join), Burlet and
Fonlupt~\cite{burlet.fonlupt:meynieltop} (amalgams), Cornu\'ejols and
Cunningham~\cite{cornuejols.cunningham:2join} (proper 2-cutset).

Note that in~\cite{cornuejols.cunningham:2join}, an operation more
general than gluing along a proper 2-cutset is considered (the
so-called \emph{2-join}, not worth defining here).  Note also that
with our definitions, it could be that for a graph $G$ with a proper
2-cutset, the blocks of decompositions are not perfect. This happens
for instance with the chordless cycle $v_1\dots v_6 v_1$ and the
proper 2-cutset $v_1, v_4$.  The blocks of decomposition are then both
isomorphic to a cycle of length~5, a notoriously non-perfect graph.
But the converse works smoothly: if two graphs are perfect, then a
perfect graph is obtained by gluing them along a proper 2-cutset.  A
proof of this is implicit in~\cite{cornuejols.cunningham:2join}.
Another simple way to check this is to note that when a vertex $v$ has
degree~2 and non-adjacent neighbors $a, b$ in a perfect graph $G$,
then all paths from $a$ to $b$ in $G$ have even length (otherwise, $G$
contains an odd chordless cycle of length at least~5).  Such a pair
$a, b$ is what is called an \emph{even pair}, and it is proved
in~\cite{fonlupt.uhry:82} that there exists an optimal coloring of $G$
such that $a$ and $b$ have the same color.  The perfection of a graph
obtained from two perfect graphs by gluing $G_1$ and $G_2$ along a
proper 2-cutset $\{a, b\}$ is then easy to prove by a direct coloring
argument: use colorings of $G_1$ and $G_2$ that both give the same
color to $a$ and $b$.

\begin{theorem}
  \label{th:pop}
  Perfect graphs are closed under the following operations: gluing
  along a clique, substitution, 1-join composition, amalgam
  composition, gluing along a proper 2-cutset.
\end{theorem}

We now turn our attention to the preservation of $\chi$-boundedness
under the operations, but there is an important technicality.  A class
of graphs is \emph{hereditary} if it is closed under taking induced
subgraphs.  The \emph{closure} of a class $\cal B$ of graphs under a
set of graph operations is the class $\cal C$ obtained from the graphs
of $\cal B$ by perfoming the operations repeatedly and in any order. A
set of operations \emph{preserves $\chi$-boundedness} if the closure
of any hereditary $\chi$-bounded class under the set of operations is
a $\chi$-bounded class.  Of course, the function that bounds $\chi$
needs not be the same in $\cal B$ and $\cal C$, and in most cases, it
is not. This leads to a potential problem: it may happen that an
operation $O_1$ preserves $\chi$-boundedness, that another operation
$O_2$ also preserves $\chi$-boundedness, but that the set of
operations $\{O_1, O_2\}$ does not preserves $\chi$-boundedness.  This
is explained in~\cite{CPST:subst}, where an actual (but slightly
artificial) example of this phenomenon is provided.

It is very easy to prove that gluing along a clique preserves
$\chi$-boundedness.  In~\cite{CPST:subst}, it is proved that
substitution preserves $\chi$-boundedness.  In~\cite{DvorakK12}, it is
proved that 1-join composition preserves $\chi$-boundedness.  But it
is not at all easy to prove for instance that the set of operations
\{1-join composition, gluing along a clique\} preserves
$\chi$-boundedness or that amalgam composition preserves
$\chi$-boundedness.  However, these are true statements, and
corollaries of the next theorem from~\cite{penev:amalgam}.

\begin{theorem}[Penev]
  \label{th:penev}
  If a class of graphs is $\chi$-bounded, then its closure under the
  following set of operations is $\chi$-bounded: \{substitution,
  amalgam composition, gluing along a clique\}.
\end{theorem}

Also gluing along a proper 2-cutset preserves $\chi$-boundedness as
shown in~\cite{CPST:subst}.

\begin{theorem}[Chudnovsky, Penev, Scott and Trotignon]
  \label{th:cpst}
  If a class of graphs is $\chi$-bounded, then its closure under the
  operation of gluing along a proper 2-cutset is $\chi$-bounded.
\end{theorem}

Note that in Theorem~\ref{th:penev}, if the class we start with is
$\chi$-bounded by a function $f$, then the closure is $\chi$-bounded
by an exponential in $f$ (something close to $g(x) = (x f(x))^x$).  In
Theorem~\ref{th:cpst}, the situation is much better, and the resulting
function is linear in the function $f$ we start with.  The function
was even improved by Penev, Thomass\'e and Trotignon,
see~\cite{penevST:14}.  Note that in \cite{CPST:subst,penevST:14} an
operation more general than the proper 2-cutset is considered (namely,
the operation of gluing along a 2-cutset, not worth defining here).

\section{A property closed under substitutions}
\label{sec:subst}

Say that a graph $G$ has Property $P_0$ if it has no edges (such a
graph is called an \emph{independent graph}).  We now define
inductively a Property $P_k$ for all $k\geq 1$ as follows: a graph $G$
has Property $P_k$ if for every induced subgraph $G'$ of $G$ there
exists an induced subgraph $H$ of $G'$ that has Property $P_{k-1}$ and
that intersects every maximal clique of $G'$.  From the definition, it
is clear that Property $P_k$ is hereditary (if a graph has it,
then so are all its induced subgraphs). 

Graphs with Property $P_1$ are exactly the graphs $G$ such that
for every induced subgraph $H$ of $G$ there exists a stable set of $H$
that intersects every maximal clique of $H$ (where a \emph{stable set}
in a graph is a set of vertices that induces an independent graph).
Graphs satisfying Property $P_1$ are known as \emph{strongly perfect
  graphs}, see~\cite{Ravindra99} for a survey about them. They form a
(proper) subclass of perfect graphs.  To the best of our knowledge,
for $k\geq 2$, graphs with Property $P_k$ were not studied so far.

The following provides examples of graphs with Property $P_k$.

\begin{lemma}
  \label{l:amkpk}
  For all $k\geq 1$, graphs with chromatic number at most $k$ have
  Property $P_{k-1}$.
\end{lemma}

\begin{proof}
  We proceed by induction on $k$.  For $k=1$, the result is obvious.
  Suppose it holds for some fixed $k\geq 1$.  Let $G$ be a graph with
  chromatic number at most $k+1$, and $G'$ an induced subgraph of $G$.
  In $G'$, there exists an induced subgraph $H$ of chromatic number at
  most $k$ that intersects all maximal cliques of $G'$: consider for
  instance the union of the first $k$ (possibly empty) colour classes
  in a colouring of $G'$ with $k+1$ colours.  By the induction
  hypothesis, $H$ has Property $P_{k-1}$.  This proves that $G$ has
  Property $P_{k}$.
\end{proof}

The following is similar to the Lov\'asz's replication
lemma, stating that perfect graphs are closed under substitutions.

\begin{lemma}
  \label{l:pkcs}
  For all $k\geq 0$, Property $P_k$ is closed under substitution.  
\end{lemma}

\begin{proof}
  We proceed by induction on $k$.  If $k=0$, we have to prove that
  substituting an independent graph for a vertex $v$ of an independent
  graph yields an independent graph, which is obvious.  So, suppose
  $k\geq 1$ and suppose Property $P_{k-1}$ is closed under
  substitution.

  Suppose that $G$ is a graph obtained from $G_1$ by substituting
  $G_2$ for $v\in V(G_1)$ and $G_1$ and $G_2$ have Property $P_{k}$.
  We will prove that $G$ contains an induced subgraph with
  Property~$P_{k-1}$ that intersect all maximal cliques of $G$.  By
  Lemma~\ref{l:hSubst}, the same proof can be done for induced
  subgraphs $G'$ of $G$.

  A maximal clique in $G$ is either a maximal clique of $G_1$ that
  does not contain $v$ (we say that such a maximal clique has
  \emph{type non-$v$}), or is equal to $K_1 \cup K_2$ where
  $K_1\cup \{v\}$ is a maximal clique of $G_1[\{v\} \cup N(v)]$ and $K_2$ is a
  maximal clique of $G_2$ (we say that such a maximal clique has
  \emph{type $v$}).

  For $i=1, 2$, because $G_i$ has Property $P_{k}$, there exists an
  induced subgraph $H_i$ of $G_i$ that has Property $P_{k-1}$ and that
  intersects every maximal clique of $G_i$.  There are now two cases.

  If $v\in V(H_1)$, then let $H$ be the graph obtained from $H_1$ by
  substituting $H_2$ for $v$.  By the induction hypothesis, $H$ has
  Property $P_{k-1}$.  Let $K$ be a maximal clique in $G$.  If $K$ is
  of type $v$, then $K\cap V(G_2)$ is a maximal clique of $G_2$, and
  it is intersected by $V(H_2)$, so it is intersected by $H$.  If $K$
  is of type non-$v$, then $K$ is a maximum clique of $G_1$, so it
  must intersect $H_1$, and not in $v$, so it intersects $H$.  We
  proved that $H$ intersects all maximal cliques of $G$.  

  If $v\notin V(H_1)$, then we set $H=H_1$.  Let $K$ be a maximal
  clique in $G$.  If $K$ is of type $v$, then
  $(K\cap V(G_1)) \cup \{v\}$ is a maximal clique of $G_1$, and it is
  intersected by $V(H_1) = V(H)$.  If $K$ is of type non-$v$, then $K$
  is a maximum clique of $G_1$, so it must intersect $V(H_1) = V(H)$.
  We proved again that $H$ intersects all maximal cliques of $G$.
\end{proof}

Since substitution is one of the simplest operation that preserves
perfection, it is worth asking whether Property $P_k$ is closed under
gluing along a clique (another simple operation that preserves
perfection and $\chi$-boundedness).  It turns out that it is not the
case for $k=1$ (examples are provided in~\cite{berge.d:sp}).  Here we
give another example showing that $P_2$ is not closed under gluing
along a clique.

\begin{figure}
\begin{center}
  \begin{tabular}{ccc}
  \parbox[b]{1.5cm}{\includegraphics[width=1cm]{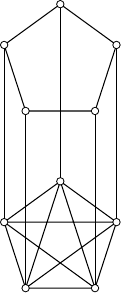}}&
  \parbox[b]{5cm}{\includegraphics[width=5cm]{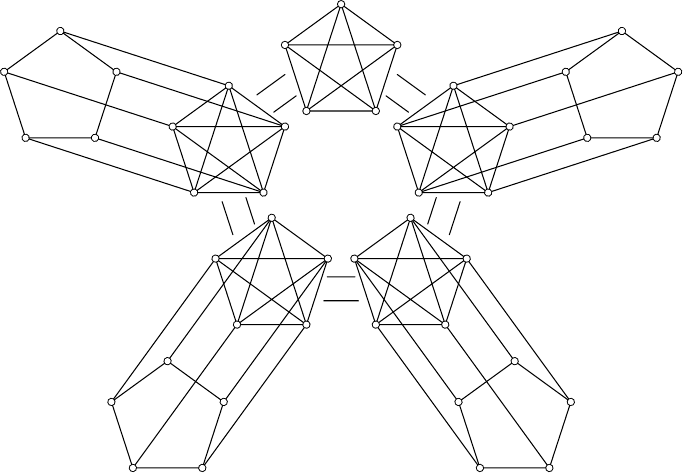}}&
  \parbox[b]{5cm}{\includegraphics[width=5cm]{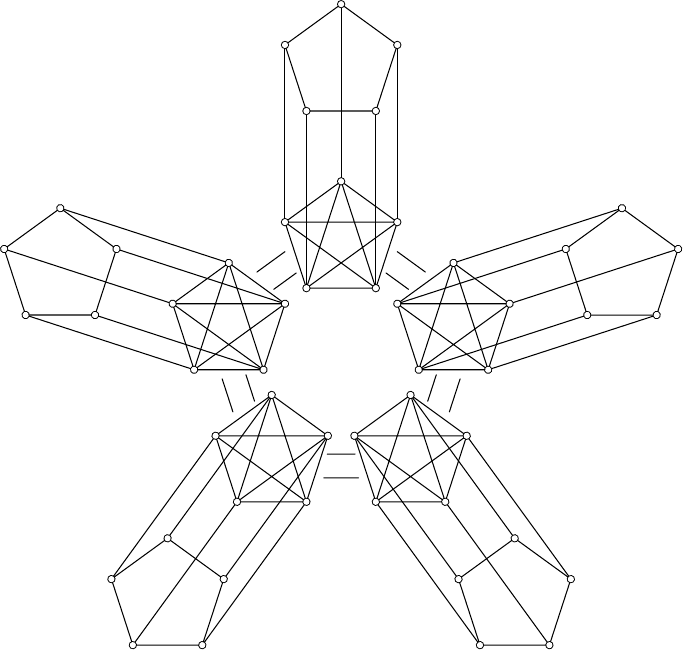}}\\
    $G_1$&$G_2$&$G_3$
  \end{tabular}
\end{center}
\caption{Three graphs\label{fig:P2}}
\end{figure}

To check this, it is convenient to rephrase Property $P_2$: for every
induced subgraph $G'$, there is a strongly perfect graph $H$ that is
an induced subgraph of $G'$ and that intersects all maximal cliques of
$G'$.  Chordal graphs are shown to be strongly perfect
in~\cite{berge.d:sp}.  On Figure~\ref{fig:P2}, three graphs are
represented.  Graph $G_1$ is obtained from a copy of $C_5$ and a copy
of $K_5$ by adding a matching.  In $G_2$, there are five copies of
$K_5$, say $H_1$, \dots, $H_5$, and there are all possible edges
between $H_i$ and $H_{i+1}$ for all $i=1, \dots, 5$ (taken modulo 5).
Four of the copies have a $C_5$ matched to them.  

It is easy to check that $G_1$ and $G_2$ both have Property $P_2$ (in
fact, they have the stronger property that a chordal graph intersects
all maximal cliques).  For instance, in $G_1$, by picking a vertex in
the $K_5$ and by taking all its non-neighbors, we obtain a chordal
graph $H$ that intersects all maximal cliques of $G_1$. In $G_2$, we
can take four copies of the chordal graph used for $G_1$ that exist in
the matched $K_5$'s.  Note that in $G_2$, no vertex of the top
clique needs to be taken in chordal graph that intersects all maximal
cliques.

However, $G_3$, that is obtained by gluing $G_1$ and $G_2$ along a
$K_5$, does not have Property $P_2$.  To see this, note that every
matching edge is a maximal clique.  Also, $H$ cannot contain a vertex
in each of the $K_5$'s (because this would form a $C_5$, that is not
strongly perfect), so at least one copy of $K_5$ does not intersect
$H$.  The $C_5$ matched to this copy therefore has to be all in $H$, a
contradiction to the strong perfection of $H$.

\vspace{2ex}

We now explain how Property $P_k$ is related to $\chi$-boundedness.
We define $f_0(0) = 0$ and $f_0(x) = 1$ for all integers $x\geq 1$.
For all integers $k\geq 1$ and $x\geq 0$, we set
$$f_k(x) =  \sum_{i=0}^{x} f_{k-1}(i).$$  By an easy induction,  for
all integer $k\geq 0$, $f_k(0)=0$, $f_k(1) = 1$ and $f_1$ is the identity
function.  Also, it is easy to check that $f_k(x) \leq x^k$ for all
integers $x, k \geq 0$.  Hence $f_k$ is a polynomial of degree $k$ (with the
convention that $0^0=0$).

\begin{lemma}
  \label{l:pkchib}
  Graphs with Property $P_k$ are $\chi$-bounded by the function $f_k$
  (and therefore by a polynomial of degree $k$).
\end{lemma}

\begin{proof}
  For $k=0$, this is trivial.  Let us prove it by induction on $k$ for
  $k\geq 1$.  Let $G$ be a graph that has Property $P_k$ and set
  $\omega = \omega(G)$.  By Property $P_k$, $G$ contains an induced
  subgraph $H_{\omega}$ that has Property $P_{k-1}$ and intersects all
  maximal cliques of $G$, so that
  $\omega(G\sm H_{\omega}) = \omega(G)-1$.  In $G\sm H_{\omega}$, 
  there exists also an induced subgraph $H_{\omega-1}$ that has
  Property $P_{k-1}$ and intersects all maximal cliques of $G\sm H_{\omega}$, and
  continuing like that, we prove that $G$ can be vertex-wise
  partitioned into $\omega(G)$ induced subgraphs $H_1$, $H_2$, \dots,
  $H_\omega$, such that for all $j=1, \dots, \omega$, $H_j$ has
  Property $P_{k-1}$ and $\omega(H_j) = j$.  By the induction
  hypothesis, we have
  $$ \chi(G) \leq \sum_{i=1}^{\omega} \chi(H_i) \leq \sum_{i=0}^{\omega}
  f_{k-1}(i) = f_k(\omega(G)).$$ The same proof can be made for all
  induced subgraphs of $G$. 
\end{proof}

\begin{theorem}
  \label{th:closK}
  The closure by substitutions of the class of $k$-colourable graphs
  is a class of graph that is $\chi$-bounded by $f_{k-1}$ (in
  particular, by a polynomial of degree $k-1$).
\end{theorem}

\begin{proof}
  Every graph in the class has Property $P_{k-1}$, either by
  Lemma~\ref{l:amkpk} or by Lemma~\ref{l:pkcs}.  So, by
  Lemma~\ref{l:pkchib}, it is $\chi$-bounded by $f_{k-1}$.
\end{proof}

As observed by Penev, for large values of $k$, a stronger result was
implicitly proved in~\cite{CPST:subst}.

\begin{theorem}[Chudnovsky, Penev, Scott and Trotignon]
  \label{th:chiSPoly}
  If a class of graphs is $\chi$-bounded by $f(x) = x^A$,
  then the closure of the class under substitution is $\chi$-bounded
  by $g(x) = x^{3A+11}$.
\end{theorem}

Since $k$-colourable graphs are $\chi$-bounded by
$f(x) = x^{\log_2 k}$, we know by Theorem~\ref{th:chiSPoly} that the
closure of $k$-colorable graphs under substitution forms a class
$\chi$-bounded by $g(x) = x^{11+3 \log_2 k}$. So, when $k$ is large,
$g(x)$ is smaller than $x^k$, but for small values,
Theorem~\ref{th:closK} provides the best bound known so far.  For
instance, the fact that the closure of 3-colorable graphs under
substitution is $\chi$-bounded by a quadratic function is seemingly a
new theorem.

\section{A property closed under amalgam and proper 2-cutset}
\label{sec:amalgam}

\newcommand{\Ki}{K^{\text{in}}}

In the rest of the paper, we adopt the unusual convention that
\emph{no vertex of a graph is complete to the empty set}.  We call
\emph{a constraint for a graph $G$} any pair $(\Ki, K^+)$ such that
$\Ki$ and $K^+$ are disjoint sets and $\Ki \cup K^+$ is a clique of
$G$.  Note that $\Ki$ and $K^+$ are therefore disjoint possibly empty
cliques of $G$.  When $(\Ki, K^+)$ is a constraint
for a graph $G$, a \emph{splitter for $(G, \Ki, K^+)$} is an induced subgraph
$H$ of $G$ that satisfies the following.

\begin{itemize}
\item $H$ intersects all maximal cliques of $G$ (except possibly $K^+$
  when $K^+$ is a maximal clique of $G$);
\item $H$ contains all vertices of $\Ki$ and $H$ contains no
  $\Ki$-complete vertex (when $\Ki = \emptyset$, this constraint can
  be forgotten);
\item $H$ contains no vertex of $K^+$.
\end{itemize}

We now define inductively a Property $Q_k$ for all $k \geq 1$. A graph
has \emph{Property $Q_1$} if it is perfect. For $k\geq 1$, a graph has
\emph{Property $Q_{k+1}$} if for every induced subgraph $G'$ and every
constraint $(\Ki, K^+)$ for $G'$, there is splitter $H$ for 
$(G', \Ki, K^+)$ with the additionnal property that $G[V(H) \cup K^+]$
has Property $Q_{k}$.  Such a splitter is called a
\emph{$k$-splitter}.

\begin{figure}
\center
{\includegraphics[width=4cm]{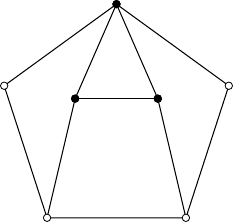}}
\caption{A graph that does not have Property~$Q_k$ for any $k$\label{f:noQk}}
\end{figure}

It is obvious that if a graph $G$ has Property $Q_k$ then every
induced subgraph of $G$ has Property $Q_k$.  On Figure~\ref{f:noQk},
we show a graph $G$ that does not have Property~$Q_k$ for any $k$.  To
see this, suppose for a contradiction that $G$ has Property~$Q_k$ for
some $k\geq 1$, and consider the minimum such $k$.  Since $G$ contains
a $C_5$ and is therefore not perfect, we have $k\geq 2$.  Define $K^+$
as the set of black vertices on the figure.  It is straightforward
that the only splitter for $(G, \emptyset, K^+)$ is $H = G \sm K^+$,
so $G[ V(H) \cup K^+] = G$ must have Property~$Q_{k-1}$, a
contradiction to the minimality of $k$.

The next lemma gives the taste of our main theorem on Property $Q_k$
(and is a particular case of it, but we prefer proving it separately).
It is very easy, but it seems impossible to prove it formally without
an induction.

\begin{lemma}
  \label{l:du}
  Property $Q_k$ is closed under disjoint union. 
\end{lemma}

\begin{proof}
  We prove the lemma by induction on $k$.  If $k=1$, the result
  follows directly from Theorem~\ref{th:pop} (because taking the
  disjoint union means gluing along an empty clique).  So, suppose
  $k\geq 1$ and let $G$ be the disjoint union of two graphs $G_1$ and
  $G_2$ that have Property $Q_{k+1}$. Let $(\Ki, K^+)$ be a constraint
  for~$G$.  Up to symmetry, we may assume that
  $\Ki\cup K^+ \subseteq V(G_1)$, and consider a $k$-splitter $H_1$
  for $(G_1, \Ki, K^+)$.  In $G_2$, we consider a $k$-splitter for
  $(H_2, \emptyset, \emptyset)$.  It is straightforward to check that
  $H_1\cup H_2$ is a splitter for $(G, \Ki, K^+)$, and by the
  induction hypothesis, it is a $k$-splitter.
\end{proof}

\begin{lemma}
  \label{l:trivialSplitter}
  For every graph $G$ and every constraint  $(\Ki, K^+)$ for $G$ there
  exist a splitter for $(G, \Ki, K^+)$. 
\end{lemma}

\begin{proof}
  Define $H$ as the graph induced by all vertices of $G\sm K^+$ that
  are not complete to $\Ki$ (in particular, if $\Ki=\emptyset$ then
  $H = G \sm K^+$).  We claim that $H$ is a splitter for
  $(G, \Ki, K^+)$.  It intersects all maximal cliques of $G$ (except
  possibly $K^+$ when $K^+$ is a maximal clique) because any vertex in
  $\Ki$ is complete to $V(G) \sm V(H)$ (and when $\Ki=\emptyset$ and
  $K^+$ is not maximal, there must be a vertex complete to $K^+$ in
  $H$).  Obviously, $H$ contains all vertices of $\Ki$, no
  $\Ki$-complete vertex, and no vertex of $K^+$.
\end{proof}

\begin{lemma}
  \label{l:monotone}
  For all $k\geq 1$, any graph with Property~$Q_k$ has
  Property~$Q_{k+1}$.
\end{lemma}

\begin{proof}
  Let $(\Ki, K^+)$ be a constraint for a graph $G$ with
  Property~$Q_k$.  Lemma~\ref{l:trivialSplitter} provides a splitter
  for $(G, \Ki, K^+)$. This splitter has Property~$Q_{k}$ because so
  does $G$.  The same proof can be done for all induced subgraphs of
  $G$, so every induced subgraph of $G$ has a $k$-splitter.
\end{proof}

\begin{lemma}
  \label{l:smallSplitter}
  Let $(\Ki, K^+)$ be a constraint for a non-bipartite
  triangle-free connected graph $G$.  Then there exists a splitter
  $H$ for $(G, \Ki, K^+)$ such that $|V(H) \cup K^+| < |V(G)|$.
\end{lemma}

\begin{proof}
  Since $G$ is triangle-free, $|\Ki \cup K^+| \leq 2$.  Also, since
  $G$ is non-bipartite and triangle-free, $|V(G)|\geq 5$.  We claim
  that $G$ contains a vertex $v$ such that:
  \begin{itemize}
  \item $v\notin \Ki\cup K^+$;
    \item $v$ has no neighbor in $K^+$;
  \item $v$ is not adjacent to any $\Ki$-complete vertex.  
  \end{itemize}

  To prove the claim, we break into cases according to the sizes of
  $\Ki$ and $K^+$.  

  If $|\Ki| = 0$ and $|K^+| = 0$, then any vertex $v$ satisfies the
  constraint. 

  If $|\Ki| = 0$ and $|K^+| = 1$, then any vertex not in $K^+$ and
  non-adjacent to the unique vertex in $K^+$ satisfies the constraint
  (and there is such a vertex, for otherwise $G$ is bipartite). 

  If $|\Ki| = 1$ and $|K^+| = 0$, then any $\Ki$-complete vertex $v$
  satisfies the constraint (and there is such a vertex since $G$ is
  connected).

  If $|\Ki| = 1$ and $|K^+| = 1$, then any $\Ki$-complete vertex not
  in $K^+$ satisfies the contraints. Since $G$ is connected, we may
  therefore assume that $\Ki$ is made of a vertex $x$ whose only
  neighbor is the vertex $y$ from $K^+$.  If all vertices of
  $G\sm K^+$ are adjacent to $y$, then since $G$ is triangle free,
  $N(y)$ is a stable set, and $G$ is bipartite, a contradiction.  It
  follows that $v$ can be chosen among the non-neighbors of $y$.

  If $|\Ki| = 2$ and $|K^+| = 0$, then no vertex in $G$ is
  $\Ki$-complete since $G$ is triangle-free, so any vertex $v$ not in
  $\Ki$ satisfies the constraint (and there exists such a vertex since
  $G$ is not bipartite).
 
  If $|\Ki| = 0$ and $|K^+| = 2$, then any vertex not in $K^+$ and
  with no neighbor in $K^+$ satisfies the constraint, so suppose that
  no such vertex exists.  It follows that $V(G) = \{x, y\} \cup N(x)
  \cup N(y)$ where $K^+ =\{x, y\}$.  Since $G$ is triangle-free, we
  see that $x\cup N(y)$ and $y \cup N(x)$ are stable sets, so $G$ is
  bipartite, a contradiction. 

  This proves the claim.  Now define $X$ as the set of $\Ki$-complete
  vertices.  Since $G$ is triangle-free, we see that
  $X \cup K^+ \cup \{v\}$ is a stable set of $G$ (except when
  $|K^+| = 2$).  It follows that $H = G\sm (X \cup K^+ \cup \{v\})$ is
  a splitter for $(G, \Ki, K^+)$.  And because of $v$, we have
  $|V(H) \cup K^+| < |V(G)|$.
\end{proof}

\begin{lemma}
  \label{l:smallQk}
  Every triangle-free graph on $n\geq 4$ vertices has Property~$Q_{n-3}$. 
\end{lemma}

\begin{proof}
  We prove the property by induction on $n$.  It is well known that
  all graphs on at most 4 vertices are perfect (they are all chordal
  except the cycle of length~4 that is bipartite).  So by
  Lemma~\ref{l:monotone}, the property is true for $n=4$.  Suppose it
  holds for $n\geq 4$, and consider a graph $G$ on $n+1$ vertices and
  a constraint $(\Ki, K^+)$ for $G$.  By Lemma~\ref{l:monotone}, we
  may assume that $G$ is not perfect (and in particular not
  bipartite).  It is enough to find an $(n-3)$-splitter for $(G, \Ki, K^+)$,
  because the proper induced subgraphs of $G$ have Property~$Q_{n-3}$
  by the induction hypothesis.  Lemma~\ref{l:smallSplitter} provides a
  splitter $H$ such that $|V(H)\cup K^+|<|V(G)|$, so the induction
  hypothesis shows that this splitter has Property $Q_{n-3}$.
\end{proof}

\begin{theorem}
  \label{Pr:Q_k}
  Property $Q_k$ is closed under the following four operations: gluing
  along a clique, amalgam (and therefore 1-join composition),
  substitution and proper 2-cutset composition.
\end{theorem}

\begin{proof}
  We proceed by induction on $k$.  For $k=1$, the result follows
  directly from Theorem~\ref{th:pop}.  Suppose it holds for some fixed
  $k \geq 1$.  We consider a graph $G$ obtained from two graphs $G_1$
  and $G_2$ with Property $Q_{k+1}$ by one of the operations, and we
  show that for every constraint $(\Ki, K^+)$, there exists a
  $k$-splitter for $(G, \Ki, K^+)$.  Each time, the splitter is
  obtained by combining $k$-splitters of $G_1$ and $G_2$ with well
  chosen constraints (they exist by assumption), and the combination
  has Property $Q_k$ by the induction hypothesis.  Note that by
  Lemma~\ref{l:lS}, the same proof can be done for induced subgraphs
  of~$G$, so we do not need to consider induced subgraphs of $G$.  Let
  us consider the operations one by one.

  \subsubsection*{Gluing along a clique}

  We suppose that $(X_1, K, X_2)$ is a split for a clique cutset of
  $G$, so $G$ is obtained from $G_1= G[X_1 \cup K]$ and
  $G_2= G[X_2 \cup K]$ by gluing along $K$. We suppose that $G_1$ and
  $G_2$ have Property $Q_{k+1}$.

  Up to symmetry, we may assume that
  $\Ki \cup K^+ \subseteq X_1 \cup K$.  Set
  $\Ki_{1}=\Ki, K^{+}_{1}=K^{+}$ and let $H_1$ be a splitter for
  $(G_1, \Ki_1, K^+_1)$.  There are two cases.

  \noindent {\bf Case 1}: $V(H_1) \cap K \neq \emptyset$. 

  Set $\Ki_{2}=V(H_1) \cap K$ and $K^{+}_{2}=K^{+} \cap K$.  Note that
  $K^{+}_{2}$ is not a maximal clique of $G_2$ since
  $\Ki_2 \neq \emptyset$.  Let $H_2$ be a splitter for
  $(G_2, \Ki_2, K^+_2)$.  Let $H=H_1 \cup H_2$.  Since $H_2$ contains
  no $\Ki_2$-complete vertex, we have $V(H_2) \cap K = \Ki_2$.  We now
  check that $H$ is a $k$-splitter for $(G, \Ki, K^+)$.

  First, $H$ contains all vertices of $\Ki$ and $H$ contains no vertex
  of $K^+$.  Vertices of $H_1$ are not complete to $\Ki_1 = \Ki$.
  Also a vertex $v\in V(H_2)$ is not complete to $\Ki$, for otherwise,
  $v\in V(H_2)\sm K$ (because as noted already
  $V(H_2) \cap K = \Ki_2$). It follows that $\Ki \subseteq K$.  But
  then, $\Ki\subseteq \Ki_2$, so $v$ is $\Ki_2$-complete, a
  contradiction.  Hence, $H$ contains no $\Ki$-complete vertex.

  Moreover, $H$ intersects all maximal cliques of $G$ (except $K^+$
  when $K^+$ is a maximal clique of $G$ and therefore of $G_1$),
  because all such cliques are either in $G_1$ or in $G_2$.

  Since $G_1[V(H_1) \cup K^+]$ and $G_2[V(H_2) \cup K^{+}_2]$ have
  Property $Q_k$ and $G[V(H) \cup K^+]$ is obtained from these two graphs
  by gluing along  $\Ki_2 \cup K^{+}_{2}$, we know   by the induction
  hypothesis that $G[V(H) \cup K^+]$ has Property~$Q_k$.

  \noindent {\bf Case 2}: $V(H_1) \cap K = \emptyset$. 

  Note that $\Ki \subseteq X_1$.  Set $\Ki_2=\emptyset$ and
  $K^{+}_{2}=K$ and let $H_2$ be a splitter for $(G_2, \Ki_2, K^+_2)$.
  Let $H=H_1 \cup H_2$.  So $H$ contains all vertices of $\Ki$, no
  vertex of $K^+$ and no $\Ki$-complete vertex.

  Observe that $G[V(H) \cup K^+]$ is obtained from
  $G_1[V(H_1) \cup K^+]$ and $G_2[V(H_2) \cup (K^+ \cap K)]$ by gluing
  along $K^+ \cap K$.  And
  $G_2[V(H_2) \cup (K^+ \cap K)]$ has Property $Q_k$ because it is an
  induced subgraph of $G_2[V(H_2) \cup K^{+}_2]$.  So by the induction
  hypothesis, $G[V(H) \cup K^+]$ has Property $Q_k$.  It remains to
  prove that $H$ intersects all maximal cliques of $G$ (except $K^+$ 
  when $K^+$ is a maximal clique of $G$). 

  Let $K'$ be any maximal clique of $G$.  Since $K$ is a clique cutset
  of $G$, $K'$ is a maximal clique of $G_1$ or $G_2$. If
  $K'\subseteq V(G_1)$, then $K'$ is intersected by $H_1$ (and therefore
  $H$) unless $K' = K_1^+ = K^+$.  If $K'\subseteq V(G_2)$, then $K'$
  is intersected by $H_2$, unless $K' = K^+_2 = K$.  In this last
  case, $K$ is a maximal clique of $G$ (and therefore $G_1$),
  and since it is not intersected by $H_1$ (because  $V(H_1) \cap
  K = \emptyset$), it must be that $K' = K_1^+ = K^+$.  In all
  cases, $H$ intersects $K'$ except  
  when $K'=K^+$.

\subsubsection*{Amalgam}
  We suppose that $(X_1,X_2, A_1, A_2, K)$ is a split for an amalgam of
  $G$. For $i=1, 2$, the \emph{block of decomposition} $G_i$
  with respect to this split is the graph obtained from $G[X_i \cup K]$
  by adding a vertex $u_{3-i}$ complete to $A_i\cup K$, so $G$ is obtained
  from $G_1$ and $G_2$ by an amalgam composition. We suppose that $G_1$ and
  $G_2$ have Property $Q_{k+1}$.
  
  Let $(\Ki,K^+)$ be a constraint for $G$.

 \noindent {\bf Case 1} $X_1 \cup K$ does not contain $\Ki  \cup K^+$ and $X_2 \cup K$ does 
  not contain $\Ki  \cup K^+$. Since $\Ki  \cup K^+$ is a clique, $\Ki  \cup K^+$ 
  belongs to $A_1 \cup A_2 \cup K$, $(\Ki  \cup K^+) \cap A_1 \neq \emptyset$,
   $(\Ki  \cup K^+) \cap A_2 \neq \emptyset$. There are three subcases.

\noindent {\bf Case 1a}: $\Ki \cap K \neq \emptyset$.

  Set $\Ki_{1}=\Ki \cap (K \cup A_1)$ and $K^{+}_1=(K^{+} \cap (K \cup A_1)) \cup \{u_2\}$ . 
  Note that $K^{+}_{1}$ is not a maximal clique of $G_1$ since
  $\Ki \cap K \neq \emptyset$. Let $H_1$ be a splitter for
  $(G_1, \Ki_1, K^+_1)$. 
  
  Set $\Ki_{2}=\Ki \cap (K \cup A_2)$ and $K^{+}_2=(K^{+} \cap (K \cup A_2)) \cup \{u_1\}$ . 
  Note that $K^{+}_{2}$ is not a maximal clique of $G_2$ since
  $\Ki \cap K \neq \emptyset$. Let $H_2$ be a splitter for
  $(G_2, \Ki_2, K^+_2)$. 
  
  Let $H=H_1 \cup H_2$.  We now check that $H$ is a $k$-splitter for $(G, \Ki, K^+)$.
 
  It is obvious that $H$ contains all vertices of $\Ki$.
  If a vertex $v \in V(H_1)$ is complete to $\Ki$, then
  $v$ is complete to $\Ki_1$, a contradiction. This implies that $H_1$ contains 
  no $\Ki$-complete vertex, no vertex of $K^+$. Similarly, $H_2$ contains 
  no $\Ki$-complete vertex, no vertex of $K^+$.
  Hence, $H$ contains no $\Ki$-complete vertex, no vertex of $K^+$.
  
  $H$ contains $\Ki$ and $\Ki \cap K \neq \emptyset$ so $H$ 
  intersects all maximal cliques of  $A_1 \cup A_2 \cup K$. Therefore, $H$ intersects 
  all maximal cliques of $G$.
  
  $G[V(H) \cup K^+]$ is obtained from two graphs $G_1[V(H_1)  \cup K^{+}_{1}]$ 
  and $G_2[V(H_2) \cup K^{+}_{2}]$    (both graphs have Property $Q_k$) by an 
  amalgam composition, so by the induction hypothesis, $G[V(H) \cup K^+]$ has 
  Property $Q_k$.
  
\noindent {\bf Case 1b}: $\Ki \cap K = \emptyset$, $\Ki \neq \emptyset$. 

  Up to symmetry, we may assume that  $\Ki \cap A_1 \neq \emptyset$.
  
  Set $\Ki_{1}=\Ki \cap A_1$ and $K^{+}_1=(K^{+} \cap (K \cup A_1)) \cup \{u_2\}$ . 
  Note that $K^{+}_{1}$ is not a maximal clique of $G_1$ since
  $\Ki \cap A_1 \neq \emptyset$. Let $H_1$ be a splitter for
  $(G_1, \Ki_1, K^+_1)$. 
  
  Set $\Ki_{2}=(\Ki \cap A_2) \cup \{u_1\}$ and $K^{+}_2=(K^{+} \cap (K \cup A_2))$ . 
  Note that $K^{+}_{2}$ is not a maximal clique of $G_2$ since
  $u_2 \notin K^{+}_{2}$. Let $H_2$ be a splitter for
  $(G_2, \Ki_2, K^+_2)$. 
  
  Let $H=G[V(H_1)\cap (V(H_2)\setminus \{u_1\})]$.  We now check that $H$ is a $k$-splitter for $(G, \Ki, K^+)$.
  
  As in the preceding case, $H$ contains all vertices of $\Ki$, no $\Ki$-complete vertex, no vertex of $K^+$.
  
  Since $H_1$ intersects all maximal cliques of $G_1$ then $P_1=V(H_1) \cap A_1 \neq \emptyset$
  and $P_1$ must intersect all maximal cliques of $A_1$ (if not this clique combined 
  with $u_2$ and $K$ would be a maximal clique of $G_1$ that $H_1$ does not intersect, a contradiction).
  This implies that $P_1$ intersects all maximal cliques of  $A_1 \cup A_2 \cup K$. Hence, $H$ intersects 
  all maximal cliques of $G$.
  
  Because  $(\Ki  \cup K^+) \cap A_2 \neq \emptyset$, $(V(H_2) \cup K^{+}_{2}) \cap A_2 \neq \emptyset$.
  Hence, $G[V(H) \cup K^+]$ is obtained from two graphs $G_1[V(H_1)  \cup K^{+}_{1}]$ 
  and $G_2[V(H_2) \cup K^{+}_{2}]$    (both graphs have Property $Q_k$) by an amalgam composition, 
  so by the induction hypothesis, $G[V(H) \cup K^+]$ has  Property $Q_k$.
  
\noindent {\bf Case 1c}: $\Ki = \emptyset$.

  If $(A_1 \cup A_2 \cup K) \sm K^{+} \neq \emptyset$, then, we choose
  $v \in (A_1 \cup A_2 \cup K) \sm K^{+}$, set $\Ki=\{v\}$ and by the
  proof of case $1$ and case $2$ we obtain a $k$-splitter $H$ for
  $(G, \Ki, K^+)$.  So, suppose $K^{+} = A_1 \cup A_2 \cup K$. This
  means $K^+$ is a maximal clique of $G$.

  Set $\Ki_{1}=\emptyset$ and $K^{+}_1=(K^{+} \cap (K \cup A_1)) \cup \{u_2\}$ . 
  Let $H_1$ be a splitter for $(G_1, \Ki_1, K^+_1)$.
  
  Set $\Ki_{2}=\emptyset$ and $K^{+}_2=(K^{+} \cap (K \cup A_2)) \cup \{u_1\}$ . 
  Let $H_2$ be a splitter for $(G_2, \Ki_2, K^+_2)$.
  
  Let $H=H_1 \cup H_2$.  We now check that $H$ is a $k$-splitter for $(G, \Ki, K^+)$.
  
  It is obvious that $H$ contains no vertex of $K^+$ and $H$
  intersects all maximal cliques of $G$ except $K^{+}$.
  
  $G[V(H) \cup K^+]$ is obtained from two graphs $G_1[V(H_1)  \cup
  K^{+}_{1}]$ and $G_2[V(H_2) \cup K^{+}_{2}]$ (both graphs have
  Property $Q_k$) by an amalgam composition, so by the induction
  hypothesis, $G[V(H) \cup K^+]$ has Property~$Q_k$.

  \noindent{\bf Case 2} 
  We are not in Case 1, so up to symmetry, we may assume that
  $\Ki \cup K^+ \subseteq X_1 \cup K$. Set
  $\Ki_{1}=\Ki, K^{+}_{1}=K^{+}$ and let $H_1$ be a splitter for
  $(G_1, \Ki_1, K^+_1)$.  There are three cases.

 \noindent {\bf Case 2a}: $V(H_1) \cap K \neq \emptyset$. 

  Set $\Ki_{2}=V(H_1) \cap K$ and $K^{+}_{2}=K^{+} \cap K$. Note that
  $K^{+}_{2}$ is not a maximal clique of $G_2$ since
  $\Ki_2 \neq \emptyset$. Let $H_2$ be a splitter for
  $(G_2, \Ki_2, K^+_2)$. 
  Since $H_2$ contains no $\Ki_2$-complete vertex, we have
  $V(H_2) \cap K = \Ki_2$, $V(H_2) \cap A_2 = \emptyset$ and $u_1 \notin V(H_2)$.
  
   Let $H=G[(V(H_1)\sm \{u_2\}) \cup V(H_2)]$.    
   We now check that $H$ is a $k$-splitter for $(G, \Ki, K^+)$.

  It is obvious that $H$ contains all vertices of $\Ki$ and $H$ contains no vertex
  of $K^+$.  Vertices of $H_1$ are not complete to $\Ki_1 = \Ki$.
  Also a vertex $v\in V(H_2)$ is not complete to $\Ki$, for otherwise,
  $v\in V(H_2)\sm K$ (because as noted already
  $V(H_2) \cap K = \Ki_2$). It follows that $\Ki \subseteq K$.  But
  then, $V(H_1) \cap K = \Ki_1=\Ki $ since $H_1$ contains no
   $\Ki_1$-complete vertex. This implies $\Ki=\Ki_2$, so $v$ is $\Ki_2$-complete, a
  contradiction.  Hence, $H$ contains no $\Ki$-complete vertex.

  $H$ contains $V(H_1) \cap K \neq \emptyset$ so $H$ intersects 
  all maximal cliques of $A_1 \cup A_2 \cup K$. Hence, $H$ intersects 
  all maximal cliques of $G$.

  Observe that $G[V(H) \cup K^+]$ is obtained from
  $G_1[(V(H_1)\sm \{u_2\}) \cup K^+_1]$ and $G_2[V(H_2) \cup K^+_2]$ by gluing
  along $K^+_2 \cap \Ki_2$.  And
   $G_1[(V(H_1)\sm \{u_2\}) \cup K^+_1]$ has Property $Q_k$ because it is an
  induced subgraph of $G_1[V(H_1) \cup K^{+}_1]$.  So by the induction
  hypothesis, $G[V(H) \cup K^+]$ has Property $Q_k$. 
  
\noindent {\bf Case 2b}: $V(H_1) \cap K = \emptyset$ and $u_2 \notin V(H_1)$. 

  $H_1$ contains all vertices of $\Ki$ so $\Ki \cap K =\emptyset$.   
  Set $\Ki_{2}=\{u_1\}$ and $K^{+}_{2}=K^{+} \cap K$. Note that
  $K^{+}_{2}$ is not a maximal clique of $G_2$ since
  $A_2 \neq \emptyset$. Let $H_2$ be a splitter for
  $(G_2, \Ki_2, K^+_2)$.  Since $H_2$ contains no $\Ki_2$-complete vertex,
  we have $V(H_2) \cap K = \emptyset$, $V(H_2) \cap A_2 = \emptyset$.
  
  Let $H=G[V(H_1) \cup (V(H_2) \sm \{u_1\})]$.
  We now check that $H$ is a $k$-splitter for $(G, \Ki, K^+)$.
  
  It is obvious that $H$ contains all vertices of $\Ki$ and $H$ contains no vertex
  of $K^+$. $H$ contains no $\Ki$-complete vertex since vertices of $H_1$ are not
  complete to $\Ki_1 = \Ki$ and $V(H_2)$ is anticomplete to $\Ki$ (because as noted already
  $V(H_2) \cap K = \emptyset$, $V(H_2) \cap A_2 = \emptyset$ and  $\Ki \cap K =\emptyset$). 
  
  Since $H_1$ intersects all maximal cliques of $G_1$ then $P_1=V(H_1) \cap A_1 \neq \emptyset$
  and $P_1$ must intersect all maximal cliques of $A_1$ (if not this clique combined 
  with $u_2$ and $K$ would be a maximal clique of $G_1$ that $H_1$ does not intersect, a contradiction).
  This implies that $P_1$ intersects all maximal cliques of  $A_1 \cup A_2 \cup K$. Hence, $H$ intersects 
  all maximal cliques of $G$.
  
  Observe that $G[V(H) \cup K^+]$ is obtained from
  $G_1[V(H_1) \cup K^+_1]$ and $G_2[(V(H_2) \sm \{u_1\}) \cup K^+_2]$ by gluing
  along $K^+_2$ (possibly empty). And $G_2[(V(H_2) \sm \{u_1\}) \cup K^+_2]$ has 
  Property $Q_k$ because it is an induced subgraph of $G_2[V(H_2) \cup K^{+}_2]$.  So by the induction
  hypothesis, $G[V(H) \cup K^+]$ has Property $Q_k$. 
  
\noindent {\bf Case 2c} $V(H_1) \cap K = \emptyset$ and $u_2 \in V(H_1)$.

  $H_1$ contains all vertices of $\Ki$ so $\Ki \cap K =\emptyset$. Also $\Ki \not \subseteq A_1$, for otherwise 
  $u_2 \in V(H_1)$ is complete to $\Ki$.
  
  Choose any vertex $v_2 \in A_2$, set $\Ki_{2}=\{v_2\}$ and $K^{+}_{2}=(K^{+} \cap K) \cup \{u_1\}$.
  Note that $K^{+}_{2}$ is not a maximal clique of $G_2$ since
  $A_2 \neq \emptyset$. Let $H_2$ be a splitter for
  $(G_2, \Ki_2, K^+_2)$.  Since $H_2$ contains no $\Ki_2$-complete vertex,
  we have $V(H_2) \cap K = \emptyset$.
  
  Let $H=G[(V(H_1) \setminus \{u_2\}) \cup V(H_2)]$.
  We now check that $H$ is a $k$-splitter for $(G, \Ki, K^+)$.
  
  It is obvious that $H$ contains all vertices of $\Ki$ and $H$ contains no vertex
  of $K^+$.  Vertices of $H_1$ are not complete to $\Ki_1 = \Ki$.
  Also a vertex $v \in V(H_2)$ is not complete to $\Ki$, for otherwise,
  $v \in A_2$ since $\Ki \cap K =\emptyset$. But then $v$ is adjacent to a vertex of $X_1 \sm A_1$ 
  (because as noted  already $\Ki \not \subseteq A_1$), a contradiction. 
   Hence, $H$ contains no $\Ki$-complete vertex.
   
  Since $H_2$ intersects all maximal cliques of $G_2$ then $P_2=V(H_2) \cap A_2 \neq \emptyset$
  and $P_2$ must intersect all maximal cliques of $A_2$ (if not this clique combined 
  with $u_1$ and $K$ would be a maximal clique of $G_2$ that $H_2$ does not intersect, a contradiction).
  This implies that $P_2$ intersects all maximal cliques of  $A_1 \cup A_2 \cup K$. Hence, $H$ intersects 
  all maximal cliques of $G$.
  
  Consider $V(H_1) \cap A_1 = \emptyset$, $G[V(H) \cup K^+]$ is obtained from 
  two graphs $G_1[(V(H_1) \setminus \{u_2\})  \cup K^+_1]$ and $G_2[V(H_2) \cup (K^+ \cap K)]$ 
 (both graphs have Property $Q_k$) by gluing along a clique $K^+ \cap K$ (possibly empty),
  so by the induction hypothesis, $G[V(H) \cup K^+]$ has Property $Q_k$.
  In the case $V(H_1) \cap A_1 \neq \emptyset$,  $G[V(H) \cup K^+]$ is obtained 
  from two graphs $G_1[V(H_1)  \cup K^+_1]$   and $G_2[V(H_2) \cup K^{+}_2]$ 
 (both graphs have Property $Q_k$) by an amalgam composition, so by
 the induction hypothesis,
  $G[V(H) \cup K^+]$ has Property $Q_k$.
  
  We are done when $\Ki \cup K^+ \subseteq X_1 \cup K$.
  \vspace{0.2cm}

Hence, $G$ has Property $Q_{k+1}$.

\subsubsection*{Substitution}

Since the amalgam is already treated, by Lemma~\ref{l:easyS}, it is
enough to prove that Property $Q_k$ is closed under adding a universal
vertex.
Let $G$ be a graph obtained from a graph $G'$ by adding a universal
vertex $v$. We suppose that $G'$ has Property $Q_{k+1}$.
Let $(\Ki,K^+)$ be a constraint for $G$. There are two cases.

\noindent {\bf Case 1}: $v \in \Ki$.

Let $H=G[\Ki]$. So $H$ contains all vertices of $\Ki$, no
vertex of $K^+$ and no $\Ki$-complete vertex. 
Every maximal clique of $G$ contains $v$ since $v$ is a universal vertex, this means $H$ intersects all maximal cliques of $G$.
$G[V(H) \cup K^+]$ is a clique, so it has Property $Q_1$. Therefore, $G[V(H) \cup K^+]$ has Property $Q_k$. Hence, $H$ is a $k$-splitter for $(G, \Ki, K^+)$.

\noindent {\bf Case 2}: $v \notin \Ki$.

Let $H$ be a splitter for $(G', \Ki, K^+ \sm \{v\})$. 

We have $v \notin H$ so in $G$, $H$ contains all vertices of $\Ki$, no vertex of $K^+$, no $\Ki$-complete vertex.

Assume $H$ does not intersect a maximal clique $P$ ($P \neq K^+$) of $G$. It follows that $H$ does not intersect a maximal clique $P \sm \{v\}$ of $G'$ since  $v$ is a universal vertex of $G$, a contradiction.

Hence, $H$ intersects all maximal cliques of $G$ (except $K^+$ when $K^+$ is a maximal clique). 

If $v \notin K^+$ then $G[V(H) \cup K^{+}]$ has Property $Q_k$ since $G[V(H) \cup K^{+}]= G'[V(H) \cup K^{+}]$.
If $v \in K^+$ then $G[V(H) \cup K^{+}]$ is obtained from graph $G'[V(H) \cup (K^+ \setminus \{v\})]$ by adding a universal vertex $v$. So by the induction hypothesis, $G[V(H) \cup K^{+}]$ has Property $Q_k$.

Hence, $G$ has Property $Q_{k+1}$.

\subsubsection*{Proper 2-cutset}
We suppose that $(X_1, X_2, a, b)$ is a split for a proper 2-cutset.
For $i=1, 2$, the block of decomposition $G_i$ with respect to
this split is the graph obtained from $G[X_i \cup \{a, b\}]$ by adding
a vertex $x_{3-i}$ complete to $\{a, b\}$, so $G$ is obtained
from $G_1$ and $G_2$ by a proper 2-cutset composition. 
We suppose that $G_1$ and $G_2$ have Property $Q_{k+1}$.

Let $(\Ki,K^+)$ be a constraint for $G$. Up to symmetry, 
we may assume that $\Ki \cup K^+ \subseteq X_1 \cup \{a,b\}$.  Set
$\Ki_{1}=\Ki, K^{+}_{1}=K^{+}$ and let $H_1$ be a splitter for
$(G_1, \Ki_1, K^+_1)$.  There are two cases.

\noindent {\bf Case 1}: $K^+ \cap \{a, b\} \neq \emptyset$.

Because $K^+$ is a clique, we have $K^{+} \cap \{a,b\} \neq
\{a,b\}$.
Without loss of generality, we can assume
$K^{+} \cap \{a,b\} = \{a\}$. Note that $b \notin \Ki$ since
$K^+ \cup \Ki$ is a clique of $G$. It follows $\Ki \subseteq X_1$.

The set $H_1$ intersects maximal clique $\{a, x_2\}$ and
$a \notin V(H_1)$ since $H_1$ does not contain vertices of $K^+_1$. It
follows $x_2 \in V(H_1)$.

\begin{itemize}
\item If $ b \in V(H_1)$ then set $\Ki_{2}=\emptyset, K^{+}_{2}=\{a, x_1\}$ and let $H_2$ be a splitter for
$(G_2, \Ki_2, K^+_2)$.

Let $H=G[(V(H_1)\sm \{x_2\}) \cup V(H_2)]$ so $H$ intersects all maximal cliques of $G$ (except possibly $K^+$ when $K^+$ is a maximal clique) because all such cliques are either in $G_1$ or in $G_2$.
 
$H$ contains all vertices of $\Ki$ and no vertex of $K^+$. Also $b$
is not complete to $\Ki$ since $b \in V(H_1)$ and $H_2 \sm \{b\}$ is
anticomplete to $\Ki$. This implies that $H$ contains no $\Ki$-complete vertex.
 
$G[V(H) \cup K^+]$ is obtained from two graphs $G_1[V(H_1)  \cup
K^{+}_{1}]$ and $G_2[V(H_2) \cup K^+_2]$ (both graphs have Property
$Q_k$) by a proper 2-cutset composition so by the induction hypothesis, $G[V(H) \cup K^+]$ has Property $Q_k$.

\item If $ b \notin V(H_1)$ then set $\Ki_{2}=\{x_1\}, K^{+}_{2}=\{a\}$ and let $H_2$ 
be a splitter for $(G_2, \Ki_2, K^+_2)$.

Let $H=G[(V(H_1)\sm \{x_2\}) \cup (V(H_2)\sm \{x_1\})]$ so $H$ intersects all maximal cliques of $G$ (except possibly $K^+$ when $K^+$ is a maximal clique) because all such cliques are either in $G_1$ or in $G_2$.

The graph $H$ contains all vertices of $\Ki$ and no vertex of
$K^+$. Also $H_2$ contains no $\Ki_2$-complete vertex so $H_2$ does not contains $a$ or $b$. It follows $H_2$ is anticomplete to $\Ki$. Hence, $H$ contains no $\Ki$-complete vertex.

$G[V(H) \cup K^+]$ is obtained from two graphs $G_1[(V(H_1) \setminus
\{x_2\})  \cup K^{+}_{1}]$ and $G_2[(V(H_2) \setminus \{x_1\})  \cup
K^{+}_{2}]$ (both graphs have Property $Q_k$) by a gluing at clique
$\{a\}$, so by the induction hypothesis, $G[V(H) \cup K^+]$ has Property $Q_k$.

\end{itemize}
\noindent {\bf Case 2}: $K^+ \cap \{a, b\} = \emptyset$.
\begin{itemize}
\item  If $V(H_1) \cap \{a,b\} = \emptyset$ then $\Ki$ does not contains $a$ or $b$. Set $\Ki_{2}=\{x_1\}, K^{+}_{2}=\emptyset$ and let $H_2$ be a splitter for $(G_2, \Ki_2, K^+_2)$.

Let $H=G[(V(H_1)\sm \{x_2\}) \cup (V(H_2)\sm \{x_1\})]$ so $H$ intersects all maximal cliques of $G$ (except possibly $K^+$ when $K^+$ is a maximal clique) because all such cliques are either in $G_1$ or in $G_2$.

The graph $H$ contains all vertices of $\Ki$ and no vertex of
$K^+$. Also $H_2$ contains no $\Ki_2$-complete vertex so $H_2$ does not
contains $a$ or $b$. It follows $H_2$ is anticomplete to $\Ki$. Hence,
$H$ contains no $\Ki$-complete vertex.

The graph $G[V(H) \cup K^+]$ is obtained from two disjoint graphs
$G_1[(V(H_1)\setminus \{x_2\})  \cup K^{+}_1]$ and $G_2[V(H_2)\sm
\{x_1\}]$ (both graphs have Property $Q_k$) so by the induction hypothesis, $G[V(H) \cup K^+]$ has Property $Q_k$.

\item If $V(H_1) \cap \{a,b\} = \{a,b\}$ and
  $\{a,b\} \cap \Ki = \emptyset$ then set $\Ki_{2}=\{a\}$.  If
  $\{a,b\} \cap \Ki \neq \emptyset$ then set
  $\Ki_{2}=\{a,b\} \cap \Ki$ and $\Ki$ is a clique so up to symmetry, we
  may assume that $\Ki_{2}=\{a\}$.

Set $K^{+}_{2}=\emptyset$ and let $H_2$ be a splitter for $(G_2, \Ki_2, K^+_2)$. Because $H_2$ does not contain $\{a\}$-complete vertices, $x_1 \notin V(H_2)$, so $H_2$ contains $b$ since $H_2$ intersects all maximal cliques.

Let $H=G[(V(H_1)\setminus \{x_2\}) \cup (V(H_2)\setminus \{a\})]$  so $H$ intersects all maximal cliques of $G$ (except possibly $K^+$ when $K^+$ is a maximal clique) because all such cliques are either in $G_1$ or in $G_2$.

The graph $H$ contains all vertices of $\Ki$ and no vertex of $K^+$.
Also $H_2$ contains no $\Ki$-complete vertex since $H_2$ does not
contain $\{a\}$-complete vertex. Hence, $H$ contains no $\Ki$-complete
vertex.

The graph $G[V(H) \cup K^+]$ is obtained from graphs $G_1[(V(H_1)\setminus
\{x_2\})  \cup K^{+}_{1}]$ and $G_2[V(H_2)\setminus \{a\}]$ (both
graphs have Property $Q_k$) by gluing along a clique $\{b\}$, so by
the induction hypothesis, $G[V(H) \cup K^+]$ has Property $Q_k$

\item If $V(H_1) \cap \{a,b\} \neq \emptyset$ and
  $V(H_1) \cap \{a,b\} \neq \{a,b\}$ then without loss of generality,
  we can assume that $V(H_1) \cap \{a,b\} =\{a\}$. Because $H_1$
  intersects the maximal clique $\{x_2, b\}$, $x_2 \in V(H_1)$. Hence,
  $\Ki \neq \{a\}$ (since if $\Ki = \{a\}$, then $x_2 \notin V(H_1)$, a
  contradiction).

Set $\Ki_2=\{x_1\}$, $K^{+}_{2}=\{a\}$ and let $H_2$ be a splitter for $(G_2, \Ki_2, K^+_2)$. Because $H_2$ does not contain $\{x_1\}$-complete vertex, $H_2$ does not contain $a$ or $b$.

Let $H=G[(V(H_1)\setminus \{x_2\}) \cup (V(H_2)\setminus \{x_1\}]$  so $H$ intersects all maximal cliques of $G$ (except possibly $K^+$ when $K^+$ is a maximal clique) because all such cliques are either in $G_1$ or in $G_2$.

The graph $H$ contains all vertices of $\Ki$ and no vertex of
$K^+$. Also $H$ contains no $\Ki$-complete vertex since $\Ki \neq \{a\}$.

The graph $G[V(H) \cup K^+]$ is obtained from two graphs $G_1[(V(H_1)\setminus
\{x_2\})  \cup K^{+}_{1}]$ and $G_2[(V(H_2)\setminus \{x_1\}) \cup
\{a\}]$ (both graphs have Property $Q_k$) by gluing at clique
$\{a\}$, so by the induction hypothesis, $G[V(H) \cup K^+]$ has Property $Q_k$.

\end{itemize}
\end{proof}

Recall that the function $f_k$ was defined in the previous section. 

\begin{lemma}
  \label{l:Qkchib}
  For all $k\geq 1$, graphs with Property $Q_k$ are $\chi$-bounded by the function
  $f_k$.  
\end{lemma}

\begin{proof}
  The proof is similar to the proof of Lemma~\ref{l:pkchib}.  In the
  induction step, when we consider a graph with Property~$Q_{k+1}$, we use
  the constraint $(\emptyset, \emptyset)$ to find an induced subgraph
  with Property~$Q_{k}$ that intersects all maximal cliques. 
 \end{proof}

 We can now prove the following theorem that is a seemingly new and
 non-trivial result.  Note that $\chi$-boundedness of the class under
 consideration can easilly be obtained by Theorem~\ref{th:penev}, but
 this approach would only provide an exponential function, while we
 provide a polynomial.

\begin{theorem}
  \label{th:closSmall}
  The closure by the set of operations $S = \{$gluing along a clique,
  substitutions, 1-join composition, amalgam compositions, gluing
  along a proper 2-cutset$\}$ of the class of graphs of triangle-free
  graphs of order at most $k+3$ is a class of graph that is
  $\chi$-bounded by $f_{k}$ (in particular, by a polynomial of
  degree $k$).
\end{theorem}

\begin{proof}
  Every graph in the class has Property $Q_k$, either by
  Lemma~\ref{l:smallQk} or by
  Theorem~\ref{Pr:Q_k}.  So, by Lemma~\ref{l:Qkchib}, it is
  $\chi$-bounded by $f_k$.
\end{proof}

\section{Structure of long-unichord-free graphs}
\label{sec:struc}

Recall that a \emph{long-unichord} in a graph is an edge that is the
unique chord of some cycle of length at least~5. A graph is
\emph{long-unichord-free} if it does not contain any long-unichord.
In this section, we prove a decomposition theorem for
long-unichord-free graphs.  We obtain its proof somehow for free, by
combining two known theorems.

The theorem below is proven in \cite{conforti.c.k.v:capfree}.  The
original statement is slightly more precise, but this one is enough
for our purpose.  A \emph{cap} in a graph is a cycle of length at
least~5, with a unique chord $ab$ such that $a$ and $b$ are at
distance two along the cycle.

\begin{theorem}[Conforti, Cornu\'ejols, Kapoor and Vu\v skovi\'c]
  \label{th:cf}
  If $G$ is a connected cap-free graph, then either:
  \begin{itemize}
  \item $G$ is chordal;
  \item $G$ is triangle-free;
  \item $G$ has a universal vertex;
  \item $G$ has a cutvertex;
  \item $G$ has an amalgam. 
  \end{itemize}
\end{theorem}

Here is useful corollary. 

\begin{theorem}
  \label{th:weakDec}
  If $G$ is long-unichord-free, then either:
  \begin{itemize}
  \item $G$ is chordal;
  \item $G$ is unichord-free; 
  \item $G$ has a universal vertex;
  \item $G$ has a cutvertex; 
  \item $G$ has an amalgam. 
  \end{itemize}
\end{theorem}

\begin{proof}
  Since a cap has a long unichord, $G$ is cap-free.  If $G$ contains a
  triangle, one of the outcome follows from Theorem~\ref{th:cf}.  And
  if $G$ is triangle-free, then every unichord of $G$ is a
  long-unichord.  So $G$ is unichord-free.
\end{proof}

\begin{figure}
  \begin{center}
    \hfill
    \parbox[c]{5cm}{\includegraphics[width=3cm]{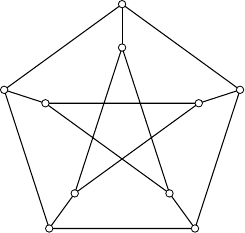}}
    \hspace{1em}
    \parbox[c]{5cm}{\includegraphics[width=3cm]{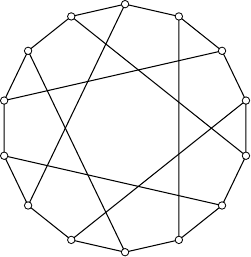}}
    \hfill
    \caption{The Petersen and the Heawood graph\label{fig:ph}}
  \end{center}
\end{figure}

The Petersen and the Heawood graphs are the graphs represented on
Figure~\ref{fig:ph}.  The following theorem is proved in \cite{nicolas.kristina:one}.

\begin{theorem}[Trotignon and Vu\v skovi\'c]
  \label{th:ucf}
  If $G$ is a connected unichord-free graph, then either: 
  \begin{itemize}
  \item $G$ is a clique;
  \item $G$ is an induced subgraph of the Petersen graph; 
  \item $G$ is an induced subgraph of the Heawood graph; 
  \item $G$ is bipartite and one side of the bipartition is made of
    vertices of degree at most~2;
  \item $G$ has a cutvertex; 
  \item $G$ has a 1-join (and therefore an amalgam);
  \item $G$ has a proper 2-cutset.
  \end{itemize}
\end{theorem}

Our main decomposition theorem is the following.

\begin{theorem}
  \label{th:MainDec}
  Let $G$ be a connected long-unichord-free graph.  Then either:
  \begin{itemize}
  \item $G$ is an induced subgraph of the Petersen graph;
  \item $G$ is an induced subgraph of the Heawood graph;
  \item $G$ is chordal;
  \item $G$ is bipartite and one side of the bipartition is made of
    vertices of degree at most~2;
  \item $G$ has a universal vertex;
  \item $G$ has a  cutvertex;
  \item $G$ has an amalgam;
  \item $G$ has proper 2-cutset.
  \end{itemize}
\end{theorem}

\begin{proof}
  We apply Theorem~\ref{th:weakDec}.  So either $G$ satisfies one of
  the outcomes, or $G$ is unichord free.  In this last case, the
  result follows from Theorem~\ref{th:ucf}. 
\end{proof}

\section{$\chi$-bounding long-unichord-free graphs}
\label{sec:cluc}

Our main purpose is to prove that all long-unichord-free graphs are
$\chi$-bounded.  This is a direct consequence of
Theorems~\ref{th:weakDec}, \ref{th:penev} and the fact proved
in~\cite{nicolas.kristina:one} that unichord-free graphs are
$\chi$-bounded.  But this approach would only provide an exponential
bound (because of Theorem~\ref{th:penev}).  Here, by using Property
$Q_k$, we prove that the class is $\chi$-bounded by a polynomial.  

By Theorem~\ref{Pr:Q_k} and~\ref{th:MainDec}, to prove that
long-unichord-free graphs have Property $Q_k$, it is enough to prove
that the basic graphs from Theorem~\ref{th:MainDec} have Property
$Q_k$.  It turns out that most of these basic graphs are perfect~: chordal
graphs and bipartite graphs are perfect, and the Heawood graph is
bipartite.  So, all these graphs have Property $Q_1$, and Property
$Q_k$ for all $k\geq 1$ by Lemma~\ref{l:monotone}. So, the only
problem is the Petersen graph, but it has Property $Q_7$ by
Lemma~\ref{l:smallQk}.  Hence, we have a short proof that
long-unichord-free graphs have all Property $Q_7$.  

\begin{figure}
\center
\includegraphics{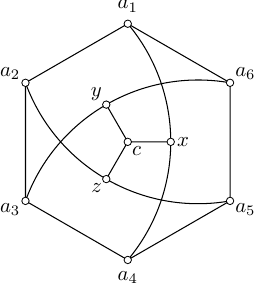}
\caption{The Petersen graph\label{fig:P6}}
\end{figure}

We now prove several lemmas needed to show that in fact,
long-unichord-free graphs have Property~$Q_3$. The only problem is to
handle the Petersen graph.  We rely on the labeling of the Petersen
graph represented on Figure~\ref{fig:P6}.  We first observe that $Q_3$
is best possible: by setting $\Ki = \{c\}$ and $K^+=\{x\}$, it can be
checked that the Petersen graph does not have Property $Q_2$.  Indeed,
the splitter $H$ cannot contain $y$ and $z$ that are $\Ki$-complete, so
it would have to contain $a_5$, $a_6$ and also $a_1$ and $a_4$
(because it does not contain $x$), so that
$H\cup \{x\}$ contains a $C_5$.

\begin{lemma}
  \label{l:spQ2}
  The graph induced by the Petersen graph on $\{a_1, a_2, a_3, a_4,
  a_5, a_6, x, c\}$ has Property~$Q_2$
\end{lemma}

\begin{proof}
  All the odd cycles of the graph under consideration go through
  $a_1$, $x$ and $a_4$, so that whatever $\Ki$ and $K^+$, one of these
  three vertices is neither in $K^+$ nor in the splitter obtained by
  taking all other vertices.  It follows that a splitter $H$ such that
  $V(H) \cup K^+$ induces a perfect graph exists. 
\end{proof}

\begin{lemma}
  \label{l:PQ3}
  The Petersen graph has Property~$Q_3$.
\end{lemma}

\begin{proof}
  We rely on the representation of the Petersen graph $G$ given on
  Figure~\ref{fig:P6}. It is well known that all pairs of adjacent
  vertices in the Petersen graph are equivalent, that is for
  $xy, x'y'\in E(G)$, there is an automorphism $\tau$ of $G$ such that
  $\tau(x) = x'$ and $\tau(y)=y'$.  This property is refered to as
  \emph{the symmetry of $G$}.  Let $G'$ be an induced subgraph of $G$,
  and $C=(\Ki, K^+)$ be a constraint for $G'$.

  If $|\Ki|=2$  (so $|K^+| = 0$), then because of the symmetry of $G$,
  we may assume that $\Ki = \{a_1, a_2\}$ and $V(G') \cap \{a_1, \dots,
  a_6, c\}$ induces a 1-splitter for $(G', \Ki, K^+)$. 

  If $|\Ki| = 1$, then we may assume because of the symmetry of $G$
  that $\Ki=\{c\}$ and $K^+ = \{x\}$ or $K^+=\emptyset$.  In both
  cases, $V(G') \cap \{a_1, \dots, a_6, c\}$ induces a 2-splitter for $(G', \Ki, K^+)$.  Note that
  $G[a_1, \dots, a_6, c, x]$ has Property~$Q_2$ by
  Lemma~\ref{l:spQ2}. 

  We may therefore assume that $\Ki= \emptyset$. Because of the
  symmetry of $G$, we may assume that $K^+ \subseteq \{x, a_1\}$.  We
  now observe as above that
  $V(G') \cap \{a_1, \dots, a_6, c, x\}\sm K^+$ induces a 1-splitter
  for $(G', \Ki, K^+)$ (it has Property~$Q_2$ as above).
\end{proof}

\begin{lemma}
  \label{l:clucFreeQ3}
  Every long-unichord-free graph has Property~$Q_3$.
\end{lemma}

\begin{proof}
  As noted at the begining of the section, a part from the Petersen
  graph, all basic graphs in Theorem~\ref{th:MainDec} are perfect (and
  have therefore Property~$Q_1$, and Property~$Q_3$ by
  Lemma~\ref{l:monotone}).  Also the Petersen graph has Property $Q_3$
  by Lemma~\ref{l:PQ3}.  The result now follows from
  Theorems~\ref{th:MainDec} and~\ref{Pr:Q_k}.
\end{proof}

\begin{theorem}
  \label{the:cluc}
  Long-unichord-free graphs are $\chi$-bounded by $f_3$ (in
  particular, by a polynomial of degree~3).
\end{theorem}

\begin{proof}
  Follows directly from Lemmas~\ref{l:clucFreeQ3} and~\ref{l:Qkchib}. 
\end{proof}

\section{Recognizing long-unichord-free graphs}
\label{sec:reco}

In this section, we describe a polytime algorithm that decides whether
a graph contains a long-unichord.  The next lemma is straightforward to
check (while a formal proof would be very long), so we prefer letting it
without proof. 

\begin{lemma}
\label{typeH}
Suppose that $(X_1, X_2, A_1, A_2, K)$ is a split of an amalgam of a
graph  $G$ and let $H$ be a hole in $G$. Then, one of the following occurs:
\begin{enumerate}
\item $V(H) \subseteq X_1$;
\item $V(H) \subseteq X_2$;
\item $H=a_1b_1a_2b_2a_1$  where $a_1,a_2 \in A_1$ and $b_1, b_2 \in A_2$;
\item $H=ba_1p_1...p_ka_2b$ where $k \geq 1$, $b \in A_2\cup K$, $a_1,a_2 \in A_1$, $p_1,...,p_k \in X_1 \setminus A_1$;
\item $H=ab_1p_1...p_kb_2a$ where $k \geq 1$, $a \in A_1\cup K$, $b_1,b_2 \in A_2$, $p_1,...,p_k \in X_2 \setminus A_2$;
\item $H=cp_1...p_kc$ where $k \geq 3$, $c \in K$, $p_1...p_k \in X_1$;
\item $H=cp_1...p_kc$ where $k \geq 3$, $c \in K$, $p_1...p_k \in X_2$;
\item $H=c_1c_2p_1...p_kc_1c_2$ where $k \geq 2$, $c_1,c_2 \in K$, $p_1...p_k \in  X_1 \setminus A_1$;
\item $H=c_1c_2p_1...p_kc_1c_2$ where $k \geq 2$, $c_1,c_2 \in K$, $p_1...p_k \in  X_2 \setminus A_2$.
\end{enumerate}
And we call them hole type $1,2,3,4,5,6,7,8,9$ respectively.
\end{lemma}
 
A \emph{decomposition} for a
graph is either an amalgam or a cutvertex, and blocks of
decomposition of these are defined in Section~\ref{sec:opkn}.

\begin{lemma}
  \label{t4}
Let $G$ be a graph that admits a decomposition. Then $G$ is a
long-unichord-free graph if and only if the blocks of decomposition
$G_1$ and $G_2$ are long-unichord-free.
\end{lemma}

\begin{proof}
  If the decomposition under consideration is a cutvertex, the result
  is clear. So, suppose it is an amalgam.  Because $G_1$ and $G_2$ are
  induced subgraph of $G$, if $G$ is long-unichord-free then both
  $G_1$ and $G_2$ are long-unichord-free.  Conversely, suppose that
  $G_1$ and $G_2$ are long-unichord-free and assume $C$ is the cycle
  with unique chord with length at least $5$ of $G$.

  Suppose first that the unique chord divides $C$ into two holes. We
  say that the type of $C$ is $XY$, according to the types $X$ and $Y$
  of the two holes with respect to the amalgam.  Based on
  Lemma~\ref{typeH}, and the property that these two holes share only
  one edges, there are $13$ possible types for $C$: $11$, $14$, $16$,
  $18$, $22$, $25$, $27$, $29$, $46$, $48$, $59$, $57$, $89$.  If $C$
  belongs to type $89$ then any vertex of $A_2$ (this vertex is exist
  since $A_2 \neq \emptyset$) with hole type $8$ induces a cycle with
  a long unichord in in $G_1$, a contradiction.  If $C$ belongs to the
  remaining types, either $G_1$ or $G_2$ contains $C$, contradiction.

  Hence, we may assume that the unique chord divides $C$ into a hole
  and a triangle ($C$ is a cap). This also means that $C$ consists of
  a hole $H$ plus a vertex $x$ that is adjacent to two adjacent
  vertices of this hole. If $H$ is of type $3$, then $x$ does not exits. If $H$
  is of type $1,2,4,5,6,7$, every choice of $x$ makes $C$ belong to $G_1$
  and $G_2$. If $H$ is of type $8$ (or $9$), this hole with a vertex of
  $A_2$ (or $A_1$) induces a cycle with a long-unichord in $G_1$, a
  contradiction.

  This proves the lemma.
\end{proof}

We now describe our algorithm (similar to the algorithm to recognize
cap-free graph from~\cite{conforti.c.k.v:capfree}).  A graph is
\emph{basic} if it is chordal or unichord-free.  
A \emph{decomposition tree} $T_G$ of a graph $G$ is defined as
follows :

\begin{itemize}
\item The root of $T_G$ is $G$. 
\item If some node $H$ of $T_G$ is not basic and has a universal
  vertex, then its unique child is $H\sm X$ where $X$ is the set of
  all universal vertices of $H$.
\item If some node of $T_G$ is not basic, has no universal vertex and
  has a decomposition, then its children are its blocks of
  decomposition.
\item All nodes not handled in the previous cases are leaves of $T_G$
  (to be more specific:  basic nodes and nodes that are not basic,
  without universal vertices and decomposition).
\end{itemize}

Note that the definition is not fully deterministic since different
decompositions can be present in a graph. In this case, one of the
decomposition should be used (so a graph may have different
decomposition trees).  Note that every graph has a decomposition tree,
and that every decomposition tree of a graph is finite (because the
children of a given node are smaller than the node).

For every graph $G$ we define
$$  f(G) = \max \left(E(\overline{G}), 1\right).$$

\begin{lemma}
  \label{l:sum}
  Suppose that $G$ is a non-basic graph, with no universal vertex and
  with an amalgam or a cutvertex. Let $G_1$ and $G_2$ be the blocks
  of decompostion of $G$ with respect to this amalgam or cutvertex.
  Then $f(G_1) + f(G_2) \leq f(G)$.
\end{lemma}

\begin{proof}
  If for $i=1, 2$ we have
  $f(G_i) = |E(\overline{G_i})$,
  then every pair $\{u, v\}$ such that $uv\notin G_i$ can be
  associated injectively to a similar pair in $G$, so the inequality
  holds.  Hence, we may assume that $f(G_1) = 1$.

  If we have
  $f(G_2) = |E(\overline{G_2})|$,
  then every pair $\{u, v\}$ such that $uv\notin G_2$ can be
  associated injectively to a similar pair in $G$.  Since $G$ has no
  universal vertex, some vertex $v \in V(G_1)\cap V(G)$ has a non-neighbor in
  $G$ and provides an extra non-adjacent pair in $G$, so that the
  inequality holds. Hence, we may assume that $f(G_2) = 1$.

  So, we just have to check that $f(G) \geq 2$. This is the case
  because $G$ is not basic, so it is not chordal and contains a
  chordless cycle of length at least~4 (that provides at least two
  non-adjacent pairs).
\end{proof}

The next lemma is implicitly proved in~\cite{ConfortiGP:15}
(as Corollary 2.16), but the machinery there is much heavier and
relies on many definitions, so we prefer to give our own simple
proof.  Note that in~\cite{cornuejols.cunningham:2join}, it is claimed
without proof that any graph with an amalgam should have an amalgam
such that at least one block of decomposition has no amalgam.  Such a
result would imply the existence of a decomposition tree of linear
size, but unfortunately, in~\cite{ConfortiGP:15}, a counter-example to
the claim is provided. 

\begin{lemma}
  \label{l:sizetree}
  Any decomposition tree of a graph $G$ has at most $O(n^2)$ nodes. 
\end{lemma}

\begin{proof}
  If a graph $H$ has a universal vertex $v$, then $f(H) = f(H\sm
  v)$. Hence, by Lemma~\ref{l:sum} the number of leaves of $T_G$
  is at most $f(G) \leq O(n^2)$.  Since removing the set of universal vertices
  can be done at most once to any node of the decomposition tree, we
  obtain the bound $O(n^2)$. 
\end{proof}

\begin{lemma}
  \label{l:DecBas}
  A graph is long-unichord-free if and only if all the leaves of its
  decomposition tree are basic. 
\end{lemma}

\begin{proof}
  Follows directly from Lemma~\ref{t4} and Theorem~\ref{th:weakDec}.
\end{proof}

\begin{theorem}
  \label{th:rec}
  Deciding whether an input graph $G$ has a long-unichord can be performed
  in time $O(n^4m^2)$ (where $n = |V(G)|$ and $m=|E(G)|$). 
\end{theorem}

\begin{proof}
  The first step is to build a decomposition tree for $G$. Deciding
  whether a graph is basic can be performed in time $O(nm)$
  (see~\cite{nicolas.kristina:one} for unichord-free graphs
  and~\cite{rose.tarjan.lueker:lbfo} for chordal graphs). Finding a
  universal vertex, a cutvertex or an amalgam can be performed in time
  $O(n^2m)$ (see~\cite{cornuejols.cunningham:2join} for the amalgam
  and the other claims are trivial).  So, the tree can be
  constructed. Once the tree is given, the algorithm checks
  whether all leaves are basics, and by Lemma~\ref{l:DecBas}, this
  decides whether the graph is long-unichord-free.

  \noindent {\bf Complexity analysis:} the most expensive step is to
  find an amalgam in time~$O(n^2m)$, and it is performed at most
  $O(n^2)$ times by Lemma~\ref{l:sizetree}.
\end{proof}

\section{Open questions}
\label{sec:open}

As observed by Esperet (personal communication), no counter-example to
the following statement is known: every hereditary $\chi$-bounded
class is $\chi$-bounded by some polynomial. This would have several
consequences. For instance, consider the following well known
conjecture:

\begin{conjecture}[Erd\H os and Hajnal, see~\cite{chudnovsky:EHconj}]
For every hereditary class $C$ of graphs, except the class of all graphs,
there exist a constant $c$ such that every graph $G$ in $C$ contains a
clique or a stable set on at least $|V(G)|^c$ vertices. 
\end{conjecture}

The conjecture is true for any class $C$ that is $\chi$-bounded by
some polynomial (because, if for some $d$, $\chi(G) \leq \omega(G)^d$,
then $\alpha(G)\omega(G)^d \geq |V(G)|$).  Let us see several results
and open problems related to the question of polynomial $\chi$-bounds.

\subsection*{``Big'' $\chi$-bounding functions}

First, let us a recall a known observation, that is seemingly
unpublished. For every integers $s, t$, define the Ramsey number
$R(s, t)$ as the smallest integer $n$ such that every graph on $n$
vertices contains a stable set of size $s$ or clique of size $t$. By
celebrated theorems of Bohman and Keevash~\cite{bohman.k.:10}, and
Ajtai, Kom{l\'o}s and Szemer{\'e}di~\cite{ajtai.k.s:80}, for every
fixed interger $s$, there exists constants $c_s, c'_s$ such that:

$$ c_s t^{s/2} \leq R(s, t) \leq c'_s t^s$$

The inequalities that we give are not as good than the ones in
papers, but are enough for our purpose. Now, for every integer $s$,
consider the class of graphs that do not contain a stable set of size
$s$ and denote it by $C_s$.  This class is clearly hereditary.  By the
definition of Ramsey numbers, a graph $G$ in $C_s$ has less than
$R(s, \omega(G)+1)$ vertices, and therefore chromatic number less than
$R(s, \omega(G)+1)$. It follows that

$$ \chi(G) \leq c'_s (\omega(G) + 1)^s$$

Hence, $C_s$ is $\chi$-bounded by some polynomial. But the interesting
point about $C_s$ is that every graph $G$  in $C_s$ has chromatic
number at least $|V(G)|/(s-1)$ (because the maximum stable set in $G$
has size at most $s-1$). Hence, if for every integer $\omega$ we
choose in $C_s$ a graph $G_{\omega}$ on $R(s, \omega+1) - 1$ vertices,
we have :  

$$  \chi(G_\omega) \geq \frac{|V(G)|}{(s-1)} 
\geq \frac{R(s, \omega+1) -
  1}{s-1}
\geq \frac{c_s}{s-1}\omega^{s/2}
$$

It follows that there cannot exist an integer $d$ such that every
hereditary class is $\chi$-bounded by a polynomial of degree $d$.  To
our knowledge, this example is the best attempt so far to construct a class
with a ``big'' $\chi$-bounding function.

\subsection*{Constraints on $\chi$-bounding functions}

More generally, what are the functions that can be the minimal
$\chi$-bounding function of some hereditary class of graph?  We
suppose that the class contains complete graphs of all sizes
(otherwise, that class is $\chi$-bounded by a constant, and any
discussion about how big can be the $\chi$-bounding function is
pointless).  Let $f$ be such a function. Let us see evidences that
there are restrictions on $f$.  Clearly, $f(0)=0$ and $f(1)=1$ and for
all integers $x$, $f(x) \geq x$. By classical constructions of
triangle-free graph with high chromatic number, we can see that $f(2)$
can be any integer higher than 1.

Suppose now that $f(2)=2$. Can $f(3)$ be any integer ? It is not the
case.  Since $f(2)=2$, we know that any graph $G$ in the class
contains no odd hole.  Therefore, by a theorem of Chudnovsly,
Robertson, Seymour and Thomas~\cite{chudnovsky.r.s.t:k4}, stating that
any odd-hole-free graph with no $K_4$ is 4-colourable, we know that
$f(3) \leq 4$.  More generally, odd-hole-free graphs are
$\chi$-bounded by a theorem of Scott and
Seymour~\cite{scottSey:oddHole}.  So, in fact, for all integers $x$,
$f(x)\leq g(x)$ where $g(x)$ is the function proven to $\chi$-bound
odd-hole-free graphs (unfortunately, this function $g$ is bigger than an
exponential).

This seems to be the only general statements that can be made about
$\chi$-bounding functions in general.  In particular, the following is
still open : suppose $f(2)=3$. Can $f(3)$ be any integer ?

\subsection*{Polynomial $\chi$-bounding functions}

For each particular $\chi$-bounded class, one might want try to prove
a polynomial bound.  The remarks above suggest that the most important
class to think of should be odd-hole-free graphs.  Also, $P_k$-free
graphs should be of interest. Because they form the simplest case
where the so-called method of extending a path developped by
Gy\'arf\'as can be applied (see~\cite{gyarfas:perfect}). And this
method seems to be the most succesfull to provide proofs of
$\chi$-boundedness, see~\cite{scottSey:oddHole,ChudnovskySS16} for
instance. However, the method notoriously produces exponential
bounds. 

The following is still open for all integers $k$ greater than 4 : is
the class of $P_k$-free graphs $\chi$-bounded by a polynomial ?

\subsection*{Polynomial $\chi$-bounds and decomposition}

It seems that the most successful attempts to prove polynomial
$\chi$-bounds make use of decomposition theorems. It is therofore
interesting to provide proofs that operations preserve the property of
being $\chi$-bounded by a polynomial.  This is known only for gluing along
a clique (trivial), substitutions (see~\cite{CPST:subst}) and gluing
along a fixed number of vertices (see~\cite{penevST:14}).  Is it true
for 1-join compositions and amalgams ?

%\bibliography{../../Bibliographie/articles}

\end{document}